\pdfoutput=1

\documentclass[11pt,twoside,a4paper,cmspaper,final,collab]{cms-tdr}
\def\svnVersion{415571}\def\cmsCernNoTag{CERN-EP/2017-049}\def\cmsCernDate{\today}\def\cmsMessage{Published in the Journal of High Energy Physics as \href{http://dx.doi.org/10.1007/JHEP07(2017)001}{\doi{10.1007/JHEP07(2017)001.}}}

\begin{document}\cmsNoteHeader{B2G-16-015}

\hyphenation{had-ron-i-za-tion}
\hyphenation{cal-or-i-me-ter}
\hyphenation{de-vices}
\RCS$Revision: 415556 $
\RCS$HeadURL: svn+ssh://svn.cern.ch/reps/tdr2/papers/B2G-16-015/trunk/B2G-16-015.tex $
\RCS$Id: B2G-16-015.tex 415556 2017-07-10 20:00:40Z jpilot $

\newcommand{\aMCATNLO} {{a\MCATNLO}\xspace}
\newcommand{\mttbar}{\ensuremath{M_{\ttbar}}\xspace}
\newcommand{\mujets}{\ensuremath{\mu\rm{+jets}}}
\newcommand{\ejets} {\ensuremath{  \rm{e+jets}}}
\newcommand{\ljets} {\ensuremath{\ell\rm{+jets}}}
\newcommand{\pp}{\ensuremath{\Pp\Pp}\xspace}
\newcommand{\W}{\ensuremath{\PW}\xspace}
\newcommand{\Wjets}{\ensuremath{\PW\text{+jets}}\xspace}
\newcommand{\Zjets}{\ensuremath{\Z\text{+jets}}\xspace}
\newcommand{\btagged} {\ensuremath{\PQb}\text{-tagged} \xspace}
\newcommand{\btagging}{\ensuremath{\PQb}\text{ tagging}\xspace}
\newcommand{\ttagged} {\ensuremath{\PQt}\text{-tagged} \xspace}
\newcommand{\ttagging}{\ensuremath{\PQt}\text{ tagging}\xspace}
\newcommand{\tmistag} {\ensuremath{\PQt}\text{ mistag} \xspace}
\newcommand{\It}  {\ensuremath{ (1\,\text{\PQt tag}) }\xspace}
\newcommand{\OtIb}{\ensuremath{ (0\,\text{\PQt tag},1\,\text{\PQb tag})}\xspace}
\newcommand{\OtOb}{\ensuremath{ (0\,\text{\PQt tag},0\,\text{\PQb tag})}\xspace}
\newcommand{\figref}[1]{Fig.~\ref{#1}}

\cmsNoteHeader{B2G-16-015}

\title{Search for \ttbar resonances in highly boosted lepton+jets and fully hadronic final states in proton-proton collisions at $\sqrt{s}=13\TeV$}

\date{\today}

\abstract{
A search for the production of heavy resonances decaying
into top quark-antiquark pairs is presented. The analysis is performed in the
lepton+jets and fully hadronic channels using
data collected in proton-proton collisions at
$\sqrt{s}=13\TeV$ using the CMS detector at the LHC, corresponding to an integrated luminosity of 2.6\fbinv. The selection is
optimized for massive resonances, where the top quarks have large
Lorentz boosts. No evidence for resonant \ttbar
production is found in the data, and upper limits on the production cross section
of heavy resonances are set. The exclusion limits for
resonances with masses above 2\TeV are significantly improved
compared to those of previous analyses at $\sqrt{s}=8\TeV$.
}

\hypersetup{%
pdfauthor={CMS Collaboration},%
pdftitle={Search for ttbar resonances in highly boosted lepton+jets and fully hadronic final states in proton-proton collisions at sqrt(s) = 13 TeV},%
pdfsubject={CMS},%
pdfkeywords={CMS, physics, B2G}}

\maketitle

\section{Introduction}
\label{sec:introduction}

Numerous extensions of the standard model (SM) predict the existence of new interactions with enhanced couplings
to third-generation quarks, especially the top quark.
The associated massive new particle contained in these theories could be observed as a \ttbar resonance in experiments at the CERN LHC.
Examples of such resonances are:
massive color-singlet
$\Z$-like bosons ($\PZpr$) in extended gauge theories~\cite{zprime_Rosner,zprime_Lynch,zprime_Carena},
colorons~\cite{Hill1991419,Jain11124928,Hill:1993hs,Hill:1994hp} and axigluons~\cite{axigluon,Choudhury:2007ux, Godbole:2008qw} in models with extended strong interaction sectors,
heavier Higgs siblings in models with extended Higgs sectors~\cite{pseudohiggs}, and
Kaluza--Klein (KK) excitations of gluons~\cite{Agashe:2006hk},
electroweak gauge bosons~\cite{Agashe:2007ki},
and gravitons~\cite{graviton}
in various extensions of the Randall--Sundrum (RS) model~\cite{RandallSundrum, Randall:1999vf}.
These models predict the existence of \TeV-scale resonances
with production cross sections of a few picobarns at $\sqrt{s} = 13$\TeV.
In all of these examples, resonant \ttbar production would be observable in
the reconstructed invariant mass spectrum of the top quark-antiquark pair ($M_{\ttbar}$).

Searches performed at the Tevatron have set upper limits
on the production cross section of
narrow $\PZpr$ resonances with masses below 900\GeV that decay into \ttbar and
have a relative decay width $\Gamma/M$ of 1.2\%~\cite{cdftt3,d0_resonance}.
Similarly, searches at the LHC
have set sub-picobarn limits
on the production cross section of resonances
in the 1--3\TeV mass range~\cite{cms-allhad,atlas-allhad, atlas-ljets,atlas-ljets-boosted,
cms-ljets,Chatrchyan:2012yca,Chatrchyan:2013lca, Aad:2015fna} at $\sqrt{s} = 7$ and 8\TeV.
The most stringent limits are from the CMS 8\TeV analysis~\cite{Khachatryan:2015sma}, which combines searches in the
fully hadronic, lepton+jets, and dilepton+jets channels. This work excludes narrow (1.2\% relative width) and wide (10\% relative width) \PZpr bosons with masses of up to 2.4 and 2.9\TeV, respectively, and an RS KK gluon with mass of up to 2.8\TeV, at the 95\% CL.

In this paper, we present a search for the production of heavy spin-1 or spin-2 resonances
decaying into \ttbar pairs using the analysis methods described in Ref.~\cite{Khachatryan:2015sma}.
We use data recorded in 2015 with the CMS detector in proton-proton ($\pp$) collisions at $\sqrt{s} = 13\TeV$  at the LHC, corresponding to an integrated luminosity of 2.6\fbinv.
Four benchmark models are considered: a $\PZpr$ boson decaying exclusively to
$\ttbar$ with relative decay widths of 1\%, 10\%, and 30\%, and a KK gluon resonance
in the RS model (having a relative decay width of approximately 17\%).
The $\PZpr$ events are generated in the framework of the sequential SM (SSM)~\cite{Altarelli:1989ff}.
Although the 1\% and 30\% widths are unphysical for various masses in that model, assuming
SM-like couplings to quarks,
this approach enables us to present limits as a function of width, allowing the results to be
reinterpreted in models with different resonance widths.
The RS KK gluon model is provided
as an example of a specific, well-motivated model with a predicted physical width.

A  search is performed using the $M_{\ttbar}$ spectrum for resonances with
masses greater than 500\GeV, where the top quarks from the resonance decay
have large Lorentz boosts.
The analysis is performed using the lepton+jets and fully hadronic \ttbar decay modes.
The lepton+jets channel is
\begin{equation*}
 \ttbar
 \to (\PWp\cPqb)(\PWm\cPaqb)
 \to {(\cPq_{1}\cPaq_{2}\cPqb) (\ell^{-} \overline{\nu}_{\ell} \cPaqb)} \qquad (\text{or charge conjugate}),
\end{equation*}
where one $\W$ boson decays hadronically, and the other decays
to a muon or an electron, and the associated neutrino.
The fully hadronic channel is
\begin{equation*}
 \ttbar
 \to (\PWp\cPqb)(\PWm\cPaqb)
 \to {(\cPq_{1}\cPaq_{2}\cPqb) (\cPq_{3}\cPaq_{4}\cPaqb)},
\end{equation*}
where both $\W$ bosons decay hadronically.
The sensitivity of the search is improved by identifying jets
originating from the hadronization of b quarks ($\cPqb$ jets),
and separating the samples into
categories that depend on the number of leptons (0 or 1), the lepton flavor (electron or muon),
the number of jets consistent with a hadronic top quark decay (``t-tagged'' jets), and the number of $\cPqb$ jets or $\cPqb$ subjets (where subjets are smaller jets found within a given jet).
In the lepton+jets channel, the resulting samples consist mainly of events from SM \ttbar production
or from  $\W$ boson production in association with jets. In the fully hadronic
channel, the resulting samples are dominated by SM \ttbar and non-top multijet production.
We refer to the latter as NTMJ, and this category comprises events from quantum chromodynamic (QCD) interactions as well as from other processes that result in jet production.  The term ``QCD multijet'' is used to describe the class of interactions considered in the generation of samples of simulated events arising solely from QCD processes.

In this paper, Section \ref{sec:detector} describes the CMS detector, while Sections \ref{sec:reconstruction} and \ref{sec:simulation} describe the techniques used for object reconstruction and the properties of simulated events utilized in the analysis, respectively.  Section \ref{sec:preselection} describes the event selections applied in each channel of the analysis, and Section \ref{sec:background} outlines the methods developed to estimate the various background components using fitting procedures.  Finally, Section \ref{sec:results} contains the results of the analysis in the form of cross section limits on new physics models, and Section \ref{sec:conclusions} summarizes the work.

\section{The CMS detector}
\label{sec:detector}

The central feature of the CMS apparatus~\cite{Chatrchyan:2008zzk}
is a superconducting solenoid of 6\unit{m} internal diameter,
providing a magnetic field of 3.8\unit{T}.
Within the solenoid volume are
a silicon pixel and strip tracker,
a lead tungstate crystal electromagnetic calorimeter (ECAL),
and a brass and scintillator hadron calorimeter (HCAL).
In the region ${\abs{\eta} < 1.74}$,
the HCAL cells have widths of 0.087 in pseudorapidity ($\eta$)
and 0.087 radians in azimuth ($\phi$).
In the $\eta$--$\phi$ plane, and for ${\abs{\eta} < 1.48}$,
the HCAL cells map on to $5{\times}5$ ECAL crystals arrays
to form calorimeter towers projecting radially outwards
from close to the nominal interaction point.
For $\abs{ \eta } > 1.74$, the coverage of the towers increases progressively to a maximum of 0.174 in $\Delta \eta$ and $\Delta \phi$. Within each tower, the energy deposits in ECAL and HCAL cells are summed to define the calorimeter tower energies, subsequently used to provide the energies and directions of hadronic jets.
Electron momenta are estimated by combining the energy measurement in the ECAL with the momentum measurement in the tracker.
Extensive forward calorimetry complements the coverage provided
by the barrel and endcap detectors.
Muons are measured in gas-ionization detectors
embedded in the steel flux-return yoke outside the solenoid.
A more detailed description of the CMS detector,
together with a definition of the coordinate system used and
the relevant kinematic variables, can be found in Ref.~\cite{Chatrchyan:2008zzk}.

\section{Event reconstruction}
\label{sec:reconstruction}

Event reconstruction is based on
the CMS particle-flow (PF) algorithm~\cite{CMS-PAS-PFT-09-001, CMS-PAS-PFT-10-001},
which takes into account information from all subdetectors, including measurements from the tracking system,
energy deposits in the ECAL and HCAL,
and tracks reconstructed in the muon detectors.
Given this information,
all particles in the event are reconstructed as
electrons, muons, photons, charged hadrons, or neutral hadrons.

Primary vertices are reconstructed using
a deterministic annealing filtering algorithm~\cite{Chatrchyan:2014fea}.
The leading primary vertex of the event is defined
as the primary vertex with the largest squared sum
of transverse momenta ($\pt$) of associated charged particles.
Charged particles associated with other primary vertices due to additional interactions within the same bunch crossing (``pileup'')
are removed from further consideration.

Muons are reconstructed
using the information collected in the muon detectors
and the inner tracking detectors, and are
measured in the range $\abs{\eta}< 2.4$.
Tracks associated with muon candidates must be consistent with
muons originating from the leading primary vertex,
and are required to satisfy
identification requirements.
Matching muon chamber information to tracks measured in the silicon tracker results
in a \pt resolution of 1.3--2.0\% in the barrel and
better than 6\% in the endcaps for muons with $20 <\pt < 100\GeV$. The \pt resolution in the barrel
is better than 10\% for muons with \pt up to 1\TeV~\cite{muonreco}.

Electron candidates are reconstructed in the range $\abs{\eta}<2.5$
by combining tracking information
with energy deposits in the ECAL.
Candidates are identified~\cite{electronreco} using information on the spatial distribution of the shower,
the track quality, and the spatial match between the track and electromagnetic cluster,
the fraction of total cluster energy in the HCAL, and
the level of activity in the surrounding tracker and calorimeter regions.
The transverse momentum resolution for electrons with $\pt\approx45$\GeV from
$\Z \to \Pe \Pe$ decays ranges from 1.7\% for nonshowering electrons
in the barrel region to 4.5\% for electrons showering in the endcaps~\cite{electronreco}.

Jets are clustered using PF candidates as inputs to the anti-\kt algorithm~\cite{Cacciari:2008gp}
in the \FASTJET~3.0 software package~\cite{FastJet}
using two different choices of the distance parameter: $R=0.4$ and 0.8.
In the following, we refer to the first set of jets
as AK4 or small-radius jets, and the second set of
jets as AK8 or large-radius jets.
For both the small- and large-radius jets, corrections based on the jet area~\cite{Cacciari:2008gn} are applied to the energy of the jets
to remove the energy contributions
from neutral hadrons from pileup interactions.
Subsequent corrections are used to account
for the combined response function of the calorimeters in both jet energy and mass,
as a function of $\eta$ and $\pt$~\cite{Chatrchyan:2011ds}.
The jet energy resolution varies from
15\% at 10\GeV to 8\% at 100\GeV to 4\% at 1\TeV for the small-radius jets, and degrades by a few percent for the large-radius jets.
The small-radius jets associated with b quarks are identified
using the Combined Secondary Vertex v2 (CSVv2) algorithm~\cite{Chatrchyan:2012jua, CMS:BTV-15-001}.
The working point used for jet $\btagging$ in this analysis
has an efficiency of $\approx$65\% (in \ttbar simulated events)
and a mistag rate (the fraction of light-flavor jets that are incorrectly tagged) of $\approx$1\%~\cite{CMS:BTV-15-001}.

The large-radius jets with $\pt > 500\GeV$ are taken as hadronic top quark candidates. To identify true top quark decays, the ``CMS top tagger v2'' algorithm~\cite{JME-15-002}
is used. In this algorithm, the constituents of the AK8 jets are reclustered using the
Cambridge--Aachen algorithm~\cite{CAcambridge,CAaachen}. The
``modified mass drop tagger'' algorithm ~\cite{mmdt}, also
known as the ``soft drop'' (SD) algorithm, recursively declusters a jet into two subjets, discarding soft and wide-angle radiation jet components until a hard splitting criterion is met, to obtain jets consistent with boosted heavy-object decays.  This algorithm has been shown to improve jet mass resolution by approximately 40\% relative to standard reconstruction techniques~\cite{CMS-JME-14-001}.  The algorithm is used with angular exponent $\beta = 0$,
soft cutoff threshold $z_{\text{cut}} < 0.1$,
and characteristic radius $R_{0} = 0.8$~\cite{Larkoski:2014wba}.
This algorithm is also able to identify two subjets within the AK8 jet.
 The subjet corresponding to the $\cPqb$ quark can be identified using subjet $\cPqb$ tagging techniques~\cite{Chatrchyan:2012jua}.
Specifically, the CSVv2 algorithm, as described above, identifies $\cPqb$-tagged subjets.  The algorithm has a comparable performance when applied to subjets, but the uncertainties are larger because of the limited number of highly boosted objects used to measure its efficiency.
The N-subjettiness observables $\tau_{\mathrm{N}}$ are calculated using all PF candidates in the AK8 jet.  Each corresponds to a $\pt$-weighted minimum distance from one of N hypothesized subjet axes, defined by the one-pass minimization procedure.  These observables are used to quantify the consistency of the particles of a jet with an N-prong decay topology.
The variable $\tau_{32} = \tau_{3}/\tau_{2}$~\cite{Thaler:2010tr, Thaler:2011gf} is employed
to identify the three-pronged substructure of a hadronically decaying top quark.
The specific working point used in this analysis is defined
by requiring that the soft-dropped mass of the jet satisfies $110<M_{\mathrm{SD}}<210\GeV$ and the N-subjettiness variable satisfies $\tau_{32}<0.69$,
which corresponds to a misidentification rate (for light-flavor quark and gluon jets) in simulation of 3\%~\cite{JME-15-002}.  This working point selects top quark jets with an efficiency of approximately 40\% when the jet $\pt$ is above 500\GeV.
Jets selected by the jet mass and N-subjettiness criteria are referred to as
``$\cPqt$-tagged''.
Additionally, $\cPqt$-tagged jets are considered to have a subjet $\cPqb$ tag if
they contain at least one soft-dropped subjet identified as $\cPqb$-tagged using the working point described above.

The missing \pt in the plane transverse to the beam direction
is reconstructed as the negative vector sum
of the \pt of all PF candidates
reconstructed in the event~\cite{Chatrchyan:2011ds}.
Its magnitude is denoted by $\ptmiss$.
Corrections to the jet energy scale and jet energy resolution
are propagated to the measurement of $\ptmiss$.

\section{Simulated events}
\label{sec:simulation}
The simulation of $\PZpr$ resonances is performed with the leading-order \MADGRAPH~v5.2.2.2~\cite{Alwall:2014hca} Monte Carlo (MC) program
using SM values for the left- and right-handed $\PZpr$ couplings to top quarks.
The simulation is performed for a range of $\PZpr$ masses between 0.5 and 4.0\TeV, and for the three relative width hypotheses
of 1\%, 10\%, and 30\%.
Higher-order QCD multijet processes
for up to three extra partons are simulated at tree level.
The $\PZpr$ boson is required to decay into
a \ttbar pair in all generated events.
The parton showering and hadronization is modeled with \PYTHIA~8.205~\cite{Sjostrand:2006za,Sjostrand:2014zea},
and the MLM algorithm~\cite{mlm} is used to match
the parton shower to the matrix element calculation
with a merging scale of $35\GeV$.

\begin{figure}[htbp]
 \centering
 \includegraphics[width=0.49\textwidth]{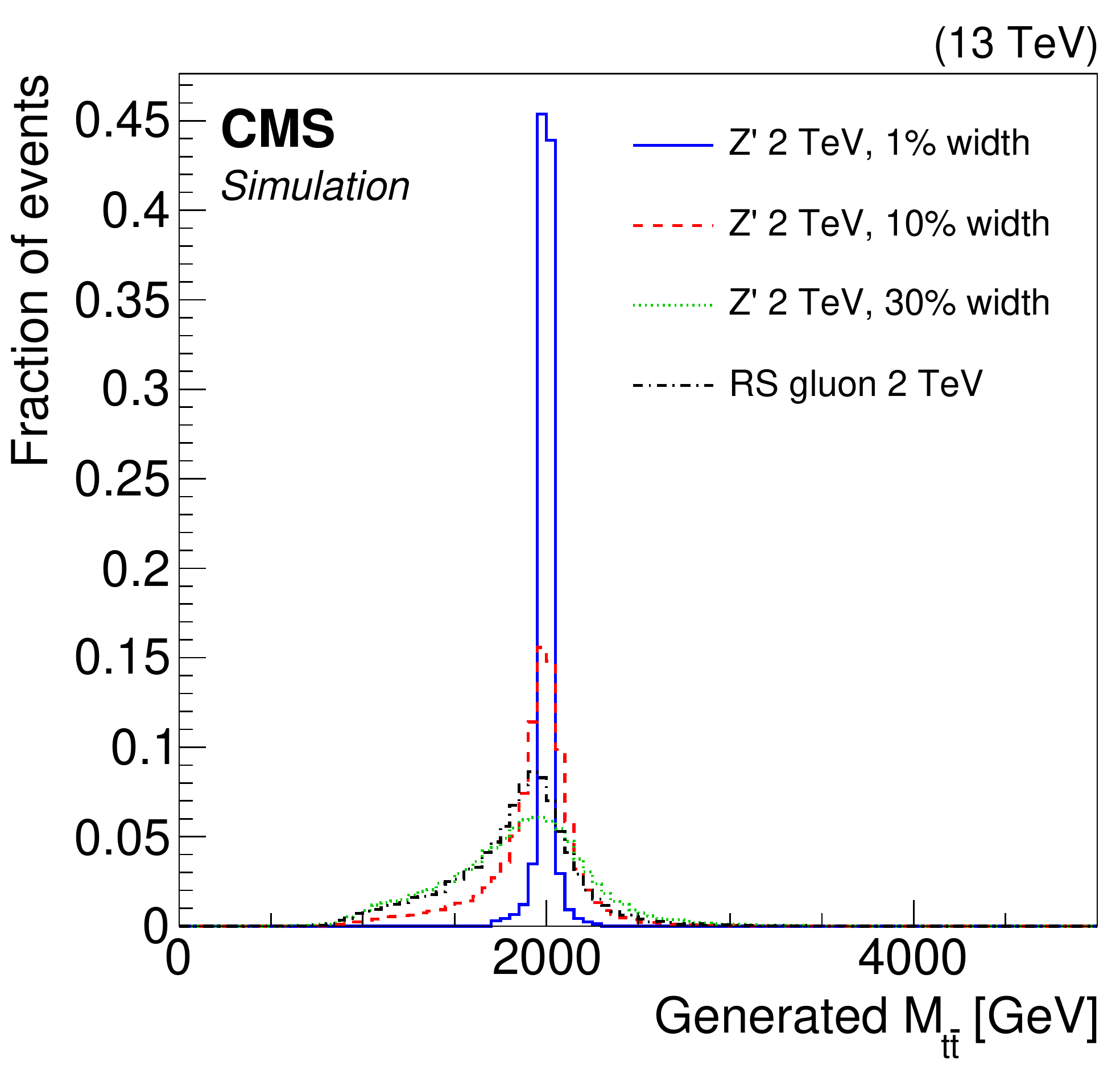}
 \includegraphics[width=0.49\textwidth]{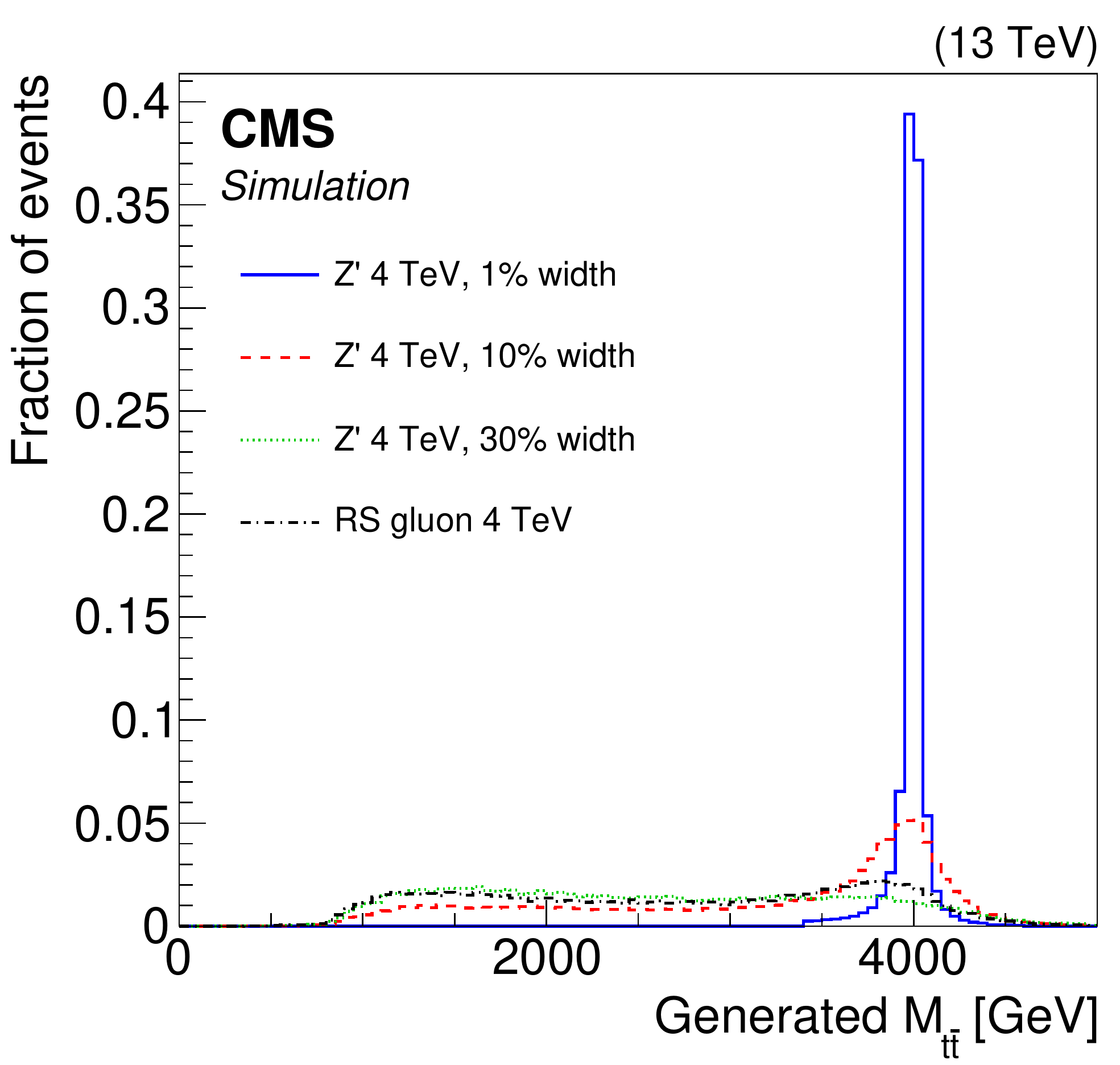}
 \caption{\label{fig:gen_mtt}
  Distributions of generator-level $M_{\ttbar}$ for the production of new particles with masses of 2\TeV (left) and 4\TeV (right), for the four signal hypotheses considered in this analysis.
  }
\end{figure}

The simulation of KK excitations of a gluon is performed with the \PYTHIA program.
The KK gluon excitations are simulated with resonance masses between 0.5 and $4.0\TeV$,  assuming the branching fraction of the KK gluon into top quark pairs is $\approx 94\%$, with the branching fraction to bottom (light) quark pairs being 5\% ($<$1\%)~\cite{Agashe:2006hk}.  Figure \ref{fig:gen_mtt} shows the generator-level $M_{\ttbar}$ distributions for resonance masses of 2\TeV and 4\TeV, for the various signal hypotheses considered.  For the highest-mass samples considered, the resonance production is dominated by off-shell contributions, giving the long tail toward low values of $M_{\ttbar}$ seen in the distributions.

Background events from \ttbar production via QCD interactions
and electroweak production of single top quarks in the tW channel
are simulated with the next-to-leading order (NLO) generator
\POWHEG (v2)~\cite{Nason:2004rx,Frixione:2007vw,Alioli:2010xd,Frixione:2007nw,Re:2010bp}.
The $s$- and $t$-channel processes of single top quark production
are simulated with \MADGRAPH{}5\_\aMCATNLO~v5.2.2.2~\cite{Alwall:2014hca}.
All events are interfaced with \PYTHIA for the description of fragmentation and hadronization.

The associated production of $\W$ or $\Z$ boson and jets is simulated using \MADGRAPH.
The MLM matching scheme is applied to match the showers generated with \PYTHIA.
Up to four additional partons in the matrix element calculations are included.
The $\ttbar$, $\W$/$\Z$+jets, and single-top-quark samples are normalized to the
theoretical predictions described in Refs.~\cite{Czakon:2011xx,
Li:2012wna, Kant:2014oha, Kidonakis:2012rm}.
Diboson processes ($\mathrm{VV} = \W\W$, $\W\Z$, and $\Z\Z$) are simulated
with \PYTHIA for both the matrix element and parton showering calculations. The event rates
are normalized to the NLO cross sections from Ref.~\cite{Campbell:2010ff}.

Simulated QCD multijet events, generated with \PYTHIA, are used to validate the
background-estimation procedure in the fully hadronic channel, but not in the search, where
the NTMJ background is estimated from sideband regions in data.

All events are generated at the center of mass
energy of $13\TeV$ and use the NNPDF~3.0 parton distribution functions (PDF)~\cite{Ball:2014uwa}.
In the parton shower simulated with \PYTHIA, the underlying event tune CUETP8M1 \cite{CMS-PAS-GEN-14-001,Skands:2014pea} has been used.
All simulated samples include the effects of additional inelastic proton-proton interactions
within the same or adjacent bunch crossings.

\section{Event selection and categorization}
\label{sec:preselection}
\subsection{Lepton+jets channel}

Events in the muon channel are collected with a single-muon trigger,
which requires the presence of a muon
with $\pt>45\GeV$ and $\abs{\eta}<2.1$.
The trigger selection employed in the electron channel
requires an electron with $\pt > 45\GeV$, $\abs{\eta}<2.5$,
and at least two jets with $\pt > 200$ (50)\GeV
for the leading (subleading) AK4 jet
reconstructed at trigger level.  These trigger choices ensure an efficiency of about 99\% for high-mass signal events.

In the lepton+jets analysis, we select events offline containing
one muon with $\pt>50\GeV$ and $\abs{\eta}<2.1$  or
one electron with $\pt>50\GeV$ and $\abs{\eta}<2.5$,
and at least two AK4 jets with $\abs{\eta} < 2.4$.
In the muon (electron) channel, the leading AK4 jet is required to have ${\pt>150}$ (${250}$)\GeV, and
the subleading AK4 jet must have ${\pt>50}$\,(${70}$)\GeV.
Additional reconstructed jets, utilized in the reconstruction of the \ttbar system,  are required to have ${\pt>30\GeV}$.
Given the highly-boosted topology of the final-state objects,
no isolation requirements are applied
to the leptons at the trigger level or in the analysis stages.
However, events are required to pass a two-dimensional selection of
$\Delta R (\ell, j) > 0.4$ or
$\pt^\text{rel}(\ell, j) > 20\GeV$,
where $j$ is the small-radius jet
with minimal angular separation $\Delta R= \sqrt{\smash[b]{(\Delta\eta)^2+(\Delta\phi)^2}}$ from the lepton $\ell$ (electron or muon),
and $\pt^\text{rel}(\ell,j)$ is the component of the
lepton momentum orthogonal to the axis of jet $j$.
The values of
$\Delta R (\ell, j)$ and
$\pt^\text{rel}(\ell, j)$
are calculated considering
small-radius jets with $\pt > 15\GeV$ and $\abs{\eta}<3.0$.
The values used are optimized for this analysis.
This two-dimensional selection effectively replaces
the more conventional lepton isolation requirement,
as it significantly reduces the background from NTMJ production
while maintaining high efficiency for the high-mass signal hypotheses.

Events in the muon channel are required to have
$\ptmiss > {50\GeV}$ and
${(\ptmiss+\pt^{\ell}) > 150\GeV}$.
In the electron channel, where jets are often misidentified as electrons,
we find that the most
effective approach for rejecting NTMJ events is to require only
${\ptmiss>120\GeV}$. After these requirements,
the contributions from NTMJ production
in both lepton channels are found to be negligible.
We also reject events that contain a second lepton
to ensure there is no overlap between the event samples and
to maintain a clear distinction between lepton+jets and dilepton+jets analyses.
Finally, we veto events with two $\ttagged$ jets to ensure
orthogonality to the fully hadronic analysis.
This veto has a negligible impact on the signal efficiency of the
lepton+jets analysis.

The kinematic reconstruction of the \ttbar system in the lepton+jets channel is performed
by assigning the products in the final state
to either the leptonic or hadronic branch of the \ttbar system.
We first assign the charged lepton and $\ptmiss$ to the leptonic branch of the event,
where $\ptmiss$ is interpreted as the $\pt$ of the neutrino, $p_z(\nu)$.
The longitudinal component of the neutrino momentum is inferred by
constraining the invariant mass of the $\ell+\nu$ system to match the W boson mass.
This procedure leads to a quadratic equation in $p_z(\nu)$.
If two real solutions are found, hypotheses are built for both cases.
If no real solutions are available, the real part is taken as $p_z(\nu)$.
In events without $\ttagged$ jets, only small-radius jets
are used to reconstruct both the leptonic and hadronic top decays.

In events containing a $\ttagged$ jet,
the large-radius jet is assigned to the hadronically decaying top quark.
Only small-radius jets with a separation of $\Delta R>1.2$ from the $\ttagged$ jet
are used in the reconstruction of the leptonic top quark decay.
Because of the presence of multiple \ttbar hypotheses per event,
a two-term $\chi^{2}$ discriminator is used to quantify
the compatibility of each hypothesis with a \ttbar decay.
The discriminator is defined as

\vspace{-0.5cm}
\begin{equation}
 \chi^{2} =   \left ( \frac{M_{\text{lep}} - \overline{M}_{\text{lep}}}{\sigma_{M_{\text{lep}}}} \right )^2
            + \left ( \vphantom{\frac{M_{lep} - \overline{M}_{lep}}{\sigma_{M_{lep}}}} \frac{M_{\text{had}} - \overline{M}_{\text{had}}}{\sigma_{M_{\text{had}}}} \right )^2,
 \label{eq:chi2}
\end{equation}
where $M_{\text{lep}}$ and $M_{\text{had}}$ are the invariant masses of
the reconstructed semileptonically and hadronically decaying top quark, respectively.
The quantities $\sigma_{M_{\text{lep}}}$ and $\sigma_{M_{\text{had}}}$ are the resolutions
of the leptonic and hadronic top quark reconstruction, respectively, and $\overline{M}_{\text{lep}}$ and $\overline{M}_{\text{had}}$ are the means of the corresponding mass distributions.
The values of
$\overline{M}_{\text{lep}}$,    $\sigma_{M_{\text{lep}}}$,
$\overline{M}_{\text{had}}$, and $\sigma_{M_{\text{had}}}$
are derived using a sample of simulated events in which all four partons of the final state top quark decay products are matched to a reconstructed jet used in the hypothesis.
In each event, the \ttbar pair reconstructed with
the smallest value of $\chi^2$ (labeled $\chi^2_\text{min}$)
is chosen.
In events with a $\ttagged$ jet, $M_{\text{had}}$ is given
by the mass of the large-radius jet calculated using the soft drop algorithm.
This choice is made because, compared to the conventional jet mass,
the soft dropped mass is much less dependent on the jet $\pt$,
and therefore on the resonance mass in a given signal hypothesis.
Moreover, this provides greater discrimination between background and signal.

Events in the signal region are required to have $\chi^2_\text{min} < 30$
for all lepton+jets categories.
This upper threshold on $\chi^2_\text{min}$
reduces the contribution of events from non-$\ttbar$ background processes
and maximizes the expected sensitivity of the analysis to new resonances.

Finally, to further enhance sensitivity, events are categorized according to
the number of $\ttagged$ and $\btagged$ jets as follows :
events with one  $\ttagged$ jet $\It$;
events with zero $\ttagged$ jets and at least one $\btagged$ jet $\OtIb$; and
events with zero $\ttagged$ and $\btagged$ jets $\OtOb$.

\subsection{Fully hadronic channel}

The fully hadronic channel requires that at least two jets satisfy
kinematic and $\cPqt$ tagging selection criteria.
The data were collected online with a trigger requiring the scalar sum of the AK4 jet energies
($H_\mathrm{T}$) to be larger than 800\GeV. The trigger selection has an efficiency of above 95\%, as measured in simulation,
for events that satisfy the offline requirement $H_\mathrm{T}> 1000\GeV$. The event reconstruction is performed using only AK8 jets.
The two leading jets are required
to have $\pt > 500$\GeV, rapidity $\abs{y} < 2.4$, and both are required to be $\cPqt$ tagged.
A back-to-back topology is selected by requiring the azimuthal separation of the two leading jets
to satisfy $\abs{\Delta \phi}> 2.1$.

Events are further categorized into six regions based on two criteria:
the rapidity difference ($\Delta y$) between the two AK8 jets and
the number of jets with at least one $\cPqb$-tagged subjet for the two highest $\pt$ jets. Events can contain 0, 1, or 2 jets with a $\cPqb$-tagged subjet, and they are separated into bins of $\abs{\Delta y} < 1.0$ and $\abs{\Delta y} > 1.0$.

\subsection{Tagging variables in lepton+jets and fully hadronic channels}

\begin{figure}[htbp]
 \centering
 \includegraphics[width=0.49\textwidth]{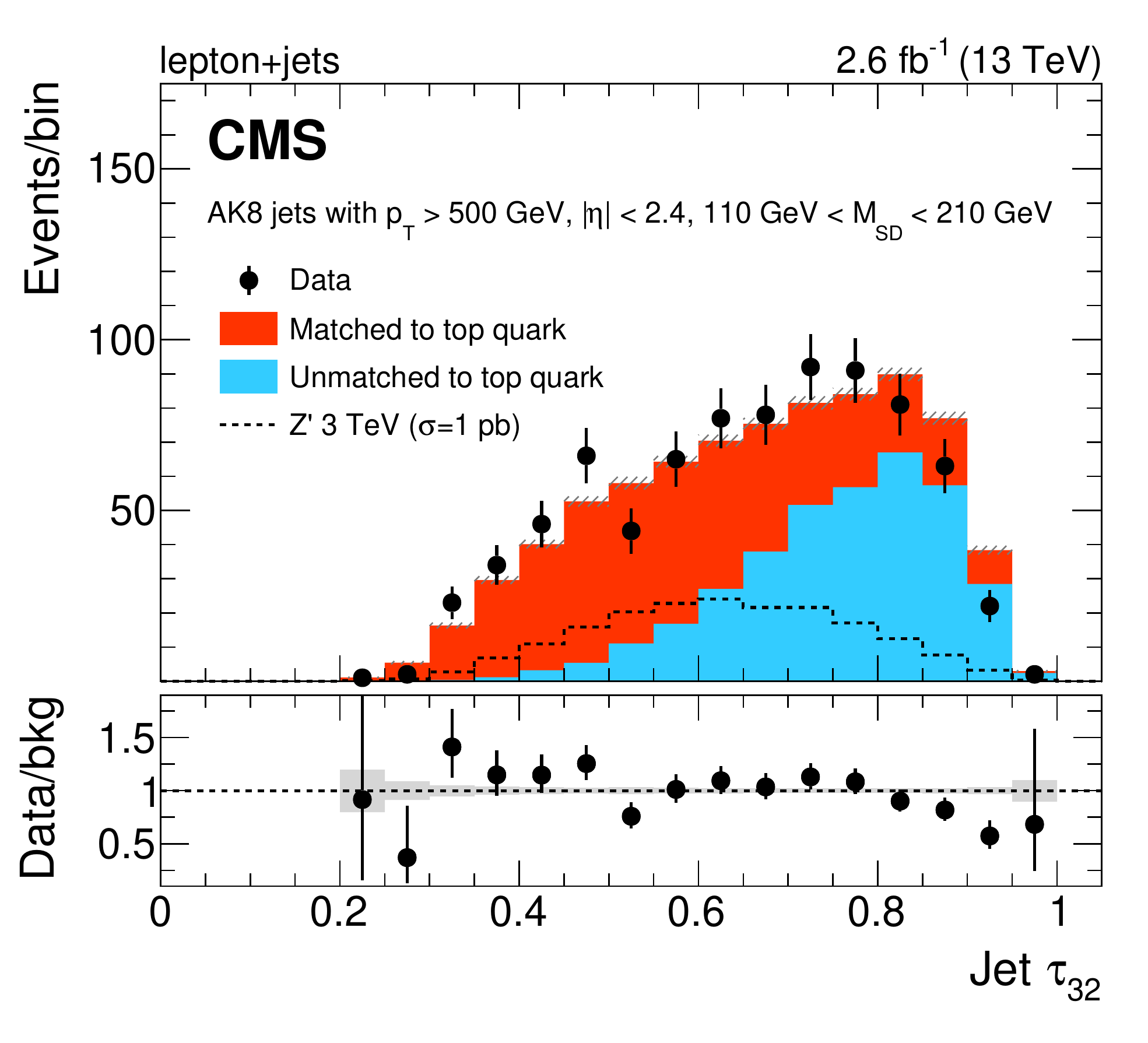}
 \includegraphics[width=0.49\textwidth]{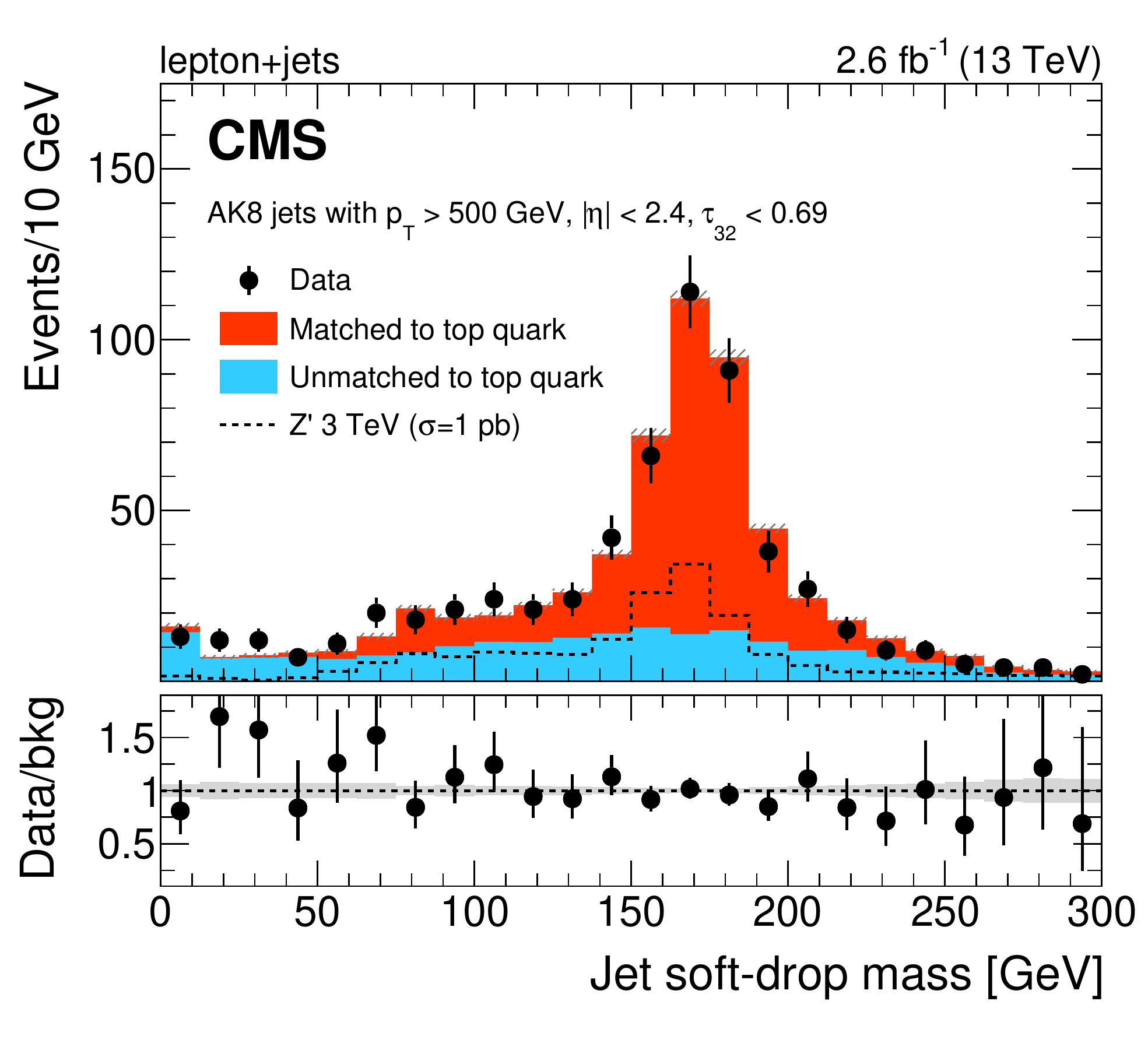}
 \includegraphics[width=0.49\textwidth]{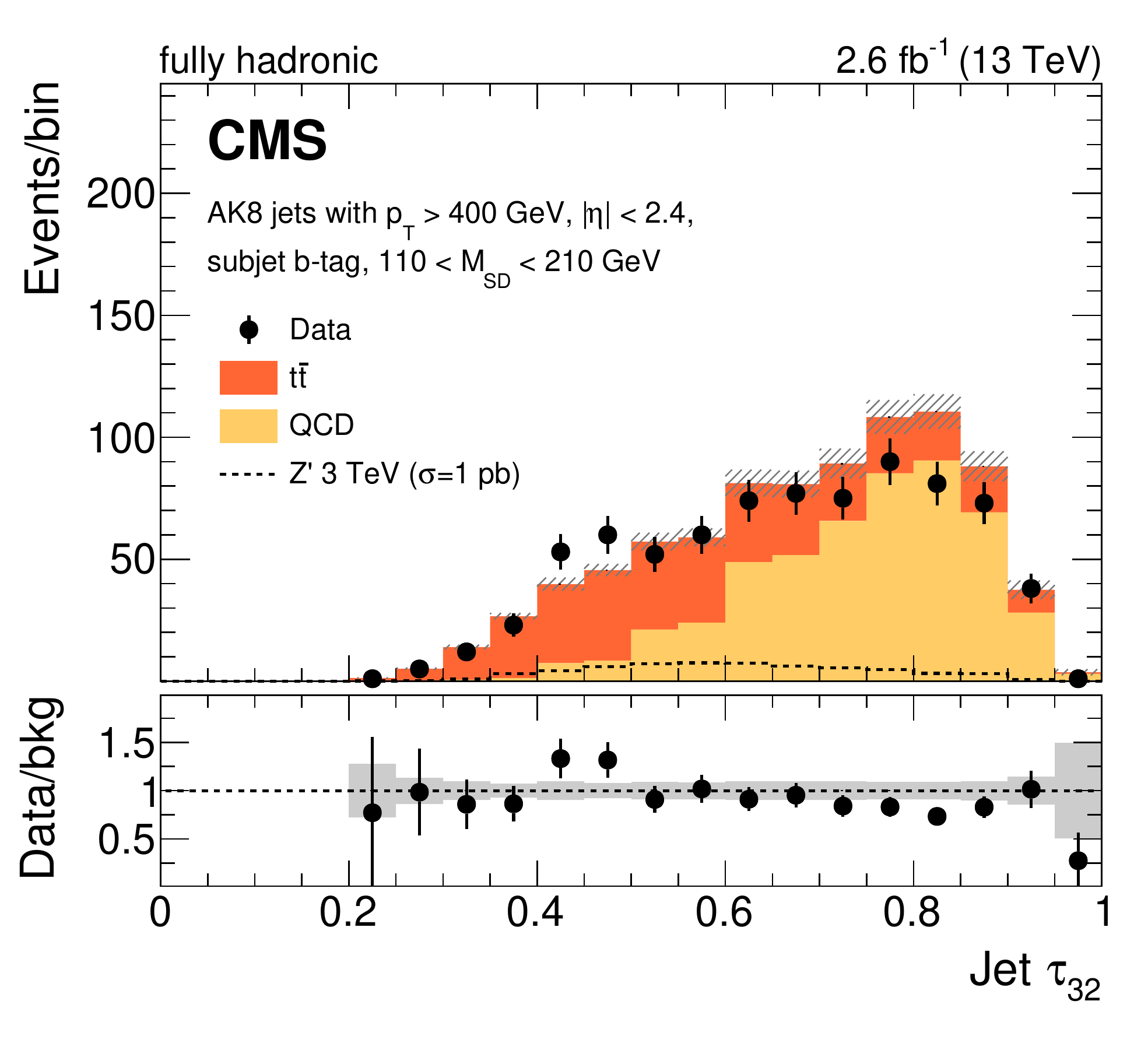}
\includegraphics[width=0.49\textwidth]{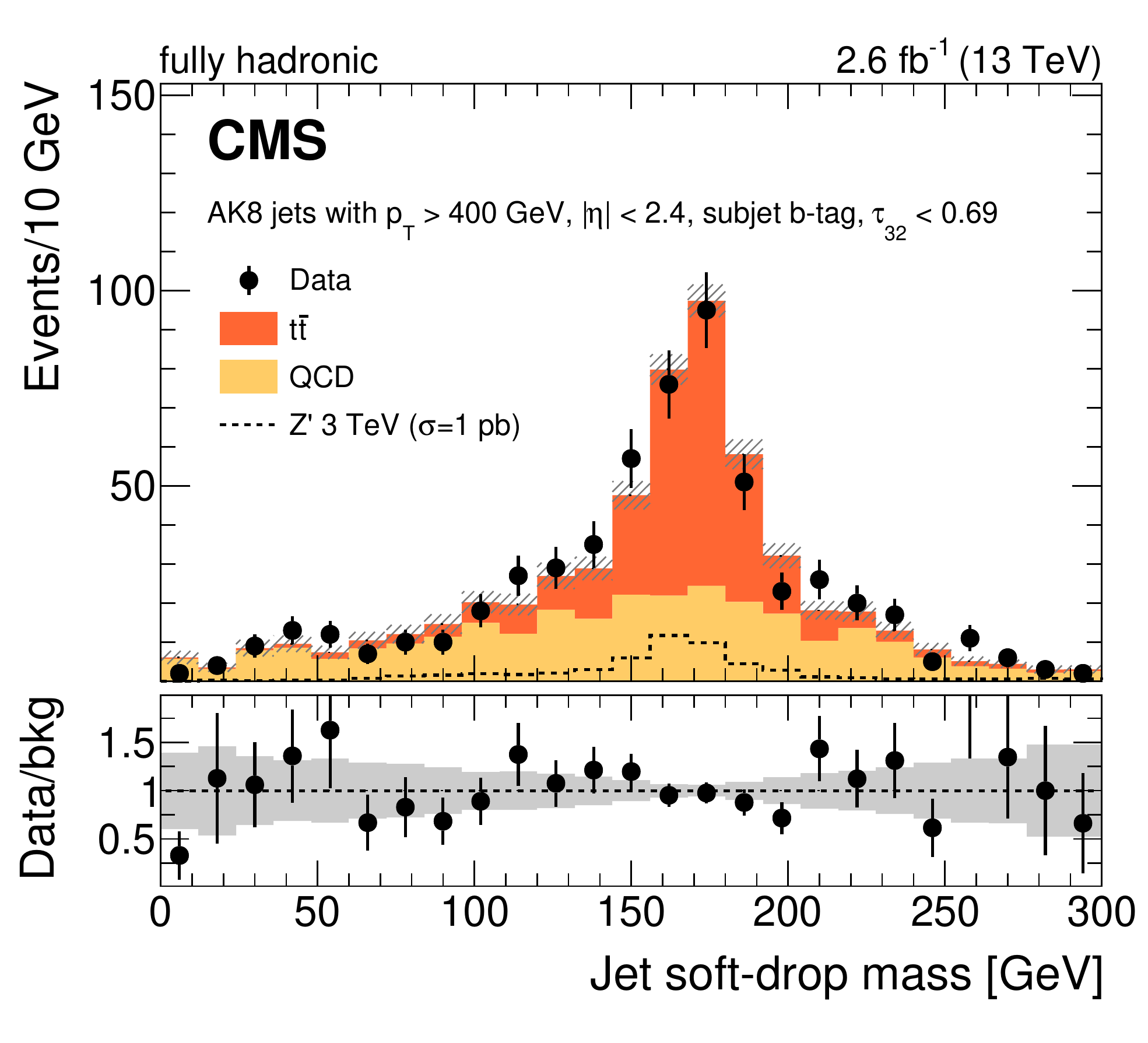}
 \caption{\label{fig:ttagging}
  Distributions of
  the N-subjettiness ratio, $\tau_{32}$, and
  the soft dropped mass, $M_{\mathrm{SD}}$,
  for AK8 jets in data and simulation, after the signal selection.  For lepton+jets, with $\pt>500\GeV$ (upper row). For the fully hadronic final state, with $\pt>400\GeV$ and subjet b tag (lower row).
  The distribution of $\tau_{32}$ (left) is shown after the selection $110<M_{\mathrm{SD}}<210\GeV$, and
  the distribution of $M_{\mathrm{SD}}$ (right) is shown after the selection $\tau_{32}<0.69$.
  The lepton+jets channel plots compare data to background simulation, where the latter is divided into contributions from jets matched at the generator level to top quarks and other jets in top pair or W+jets events. The fully hadronic channel plots compare data to \ttbar and QCD multijet simulation. Contributions from a benchmark narrow $\PZpr$ signal model are shown with the black dashed lines.  In obtaining the final results, NTMJ production is estimated from data, and simulated QCD multijet events are not used. In all plots, the error bars include only statistical contributions.
 }
\end{figure}

The distributions of the two variables used in the $\ttagging$ algorithm,
$\tau_{32}$ and $M_{\mathrm{SD}}$, are shown in Fig.~\ref{fig:ttagging} for the lepton+jets channel (upper row) and the fully hadronic channel (lower row).
Each of the figures is obtained after removing the selection on
the quantity being plotted, while maintaining all other analysis-level selections.
We observe good agreement between data and simulation in the
lepton+jets decay channel, where simulated events are divided into contributions from generator-level top quarks and other jets from \ttbar events and subdominant background processes. The fully hadronic channel also shows good agreement between the simulated distribution and data. The small discrepancies do not affect the analysis, as it relies on data to estimate the NTMJ contribution to the background.  Some discrepancy is visible at high values of $\tau_{32}$, however this region is excluded by the selection used for $\ttagging$.

\section{Background model and normalization}
\label{sec:background}
In this section, we describe the sources of the SM background and methods of background estimation for both the lepton+jets and fully hadronic channels. We then introduce the sources of systematic uncertainty considered in this analysis. Finally, we describe the treatment of the backgrounds and uncertainties in the maximum likelihood fit that is used to determine the total yield of SM processes and in the statistical analysis of data.

\subsection{Lepton+jets channel}

Several SM processes contribute
to the sample obtained from the lepton+jets selection
described in Section \ref{sec:preselection}.
The two main background processes
are \ttbar and $\Wjets$ production.
The latter accounts for a sizeable portion
of the background in the $\OtOb$ category,
whereas the former fully dominates
the $\OtIb$ and $\It$ categories.
Single top quark, $\Zjets$, and diboson production
contribute only a small fraction of the background.

The distributions obtained from simulation are
corrected to account for known discrepancies
in the observed number of data and simulated events. In particular, we derive a scale factor (SF) between data and simulation for the t tagging mistag ($\tmistag$) rate for AK8 jets from a sample dominated by $\Wjets$, selected by requiring events to have $\chi_{\text{\tiny min}}^{2}>30$. The remaining contamination from \ttbar is
  removed by subtracting the distribution of \ttbar events in simulation. The $\tmistag$ rate is measured separately for the muon and electron channels, in data and simulation. The resulting values, together with the data-to-simulation SFs, are shown in Table~\ref{tab:tmistag_effy}. As the SFs for the muon and electron channels are consistent, the weighted average is used: ${\text{SF}_{\ell} = 0.79 \pm 0.15}$.

\begin{table}[h!bt]
\centering
\topcaption{The mistag rates in data and simulation, and their ratio (data/simulation SF), for AK8 jets in the lepton+jets analysis.}
\begin{tabular}{lccc}
Channel & Eff. in data & Eff. in MC & SF                \\
\hline
e+jets   &   $0.038 \pm 0.010$               &   $0.051 \pm 0.002$             &   $0.74 \pm 0.20$ \\
$\mu$+jets   &   $0.043 \pm 0.012$               &   $0.051 \pm 0.002$             &   $0.85 \pm 0.24$ \\
\end{tabular}
\label{tab:tmistag_effy}

\end{table}

The final background estimates in this search are determined
by fitting the background-only hypothesis to data~\cite{theta}.
Distributions defined in samples dominated by various backgrounds are used
simultaneously in a binned maximum likelihood fit to constrain the different uncertainties in the background model using the data.
The reconstructed $\mttbar$ distribution is used in regions dominated by \ttbar and $\Wjets$, and the dimuon invariant mass is used in a region dominated by $\Zjets$.  The $\ttbar$-dominated region is defined by $\mttbar<2\TeV$ and $\chi_{\text{\tiny min}}^{2}<30$.
The region dominated by $\Wjets$ events is defined by
$\chi_{\text{\tiny min}}^{2}>30$.
For each of these two latter regions, six exclusive categories are defined based on lepton flavor and number of \btagged and \ttagged jets ($\It$; $\OtIb$; $\OtOb$), giving a total of 12 control regions (CRs).
One additional CR, dominated by $\Zjets$, is defined by
removing the lepton veto from the $\mujets$ selection
and adding the $\Z$ boson mass window requirement $71 <M_{\mu\mu}<111\GeV$.
The $\Z\to\Pe\Pe$ channel is
not used because of the stringent requirement on $\ptmiss$.

\subsection{Fully hadronic channel}
\label{sec:bkg_allhad}

The fully hadronic channel has two primary sources of SM background: \ttbar events and NTMJ production.
The shape of the $M_{\ttbar}$ distribution for \ttbar events is taken from simulation. The normalization of this distribution is initially set to the theoretical cross section, but is allowed to vary within both rate and shape uncertainties during the statistical analysis. The shape and normalization are both fitted and extracted for each of the six event categories. The variation of the \ttbar contribution to the total background predominantly affects the signal regions with two subjet $\cPqb$ tags, which have \ttbar as the dominant background component.

\begin{figure}[hbtp]
\centering
\includegraphics[width=0.45\linewidth]{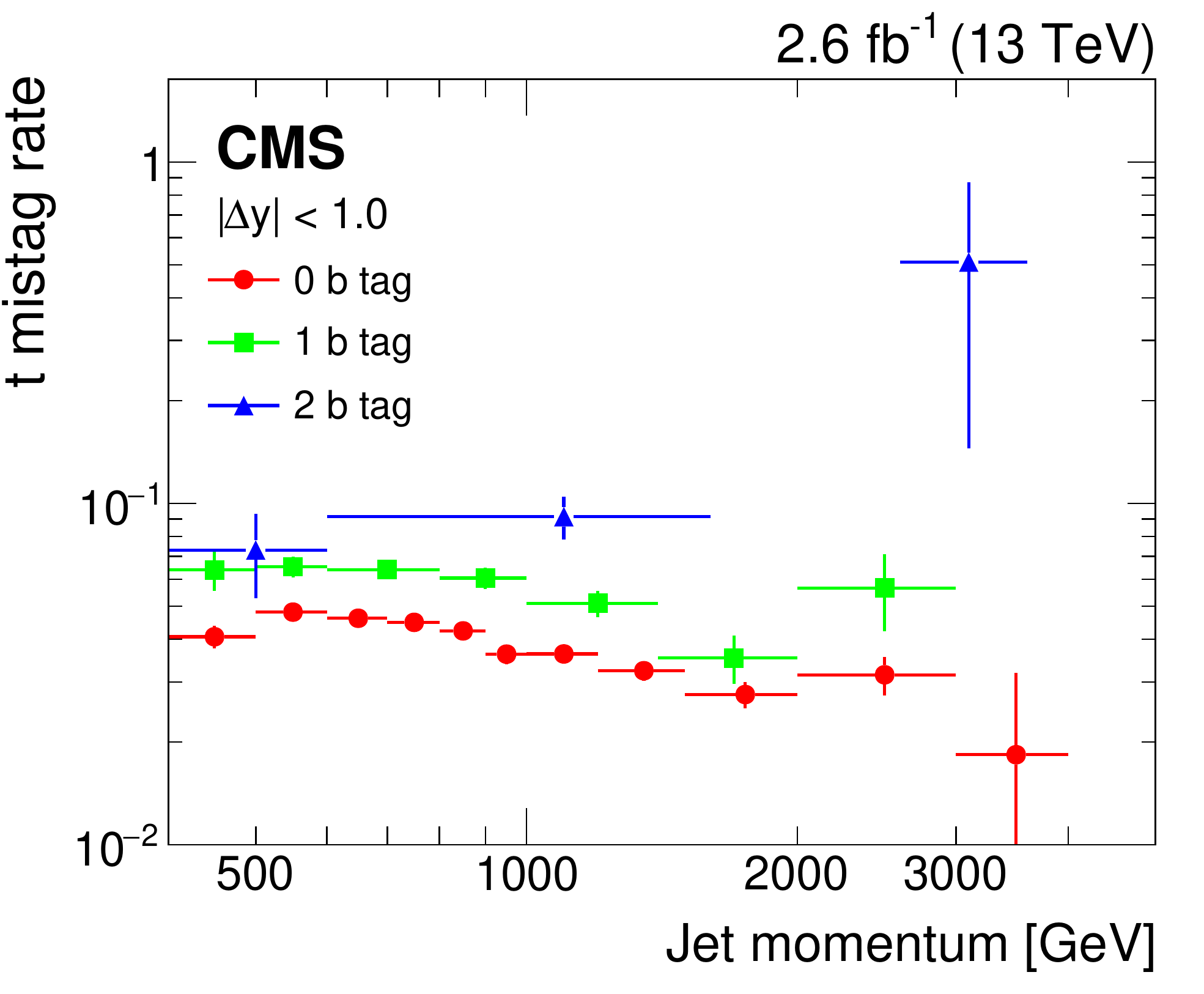}
\includegraphics[width=0.45\linewidth]{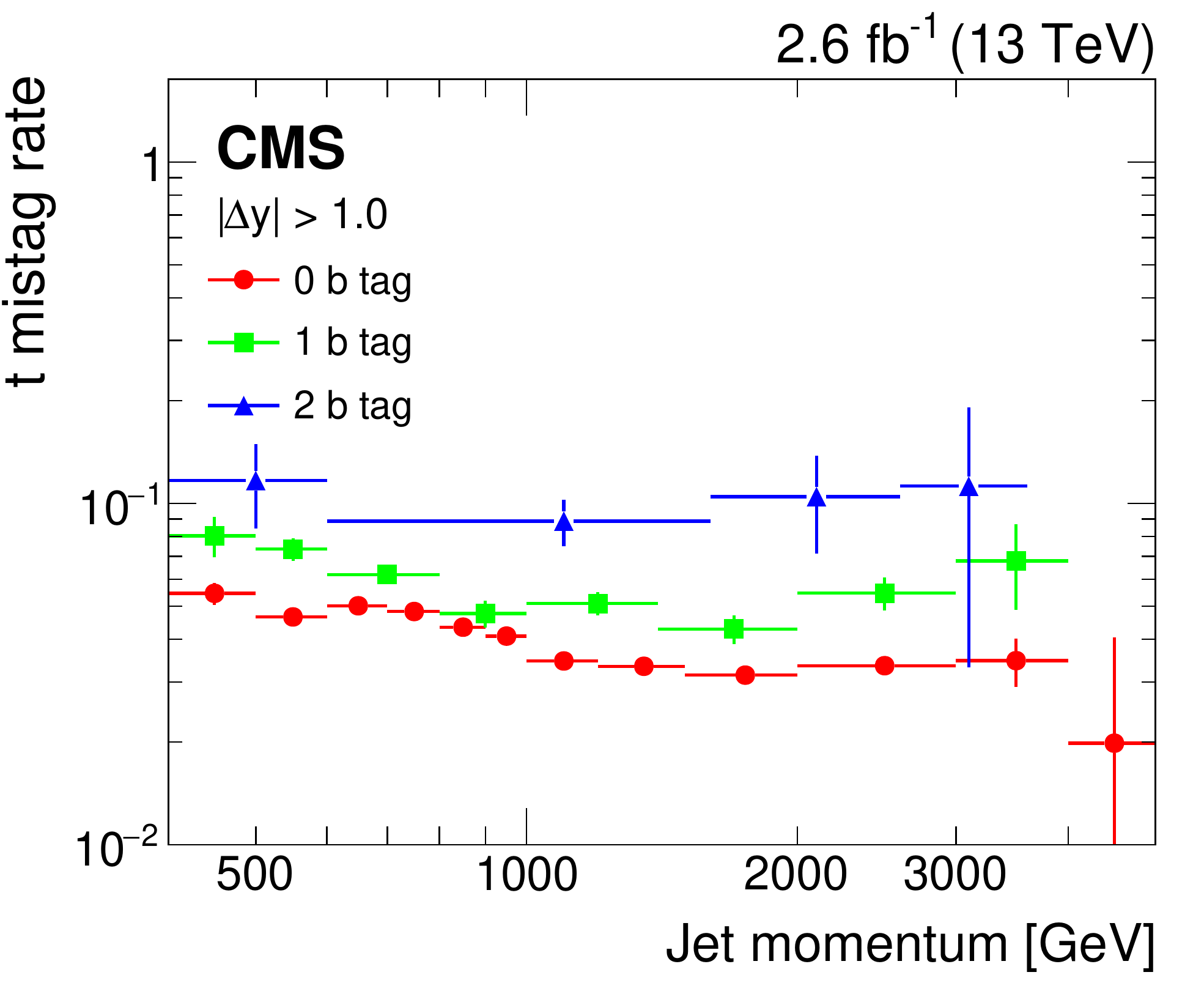}
\caption{The mistag rate for the $\cPqt$ tagging algorithm in the fully hadronic channel, measured with data for the six event categories by an anti-tag and probe
  procedure. The round, square, and triangular points
  indicate the $\tmistag$ rate for events in the 0, 1, and 2
  $\cPqb$ tag categories, respectively. The left (right) plot contains
  events with $\abs{\Delta y} < 1.0$ ($> 1.0$). The contamination
  from \ttbar production is removed by subtracting the distribution of \ttbar events in simulation,
  normalized to SM expectation.}
\label{fig:mistag}

\end{figure}

For the NTMJ estimate, we use a data-driven technique similar to that described in Ref.~\cite{Chatrchyan:2013lca}.
The method involves selecting a sample of data events with low SM \ttbar contribution by inverting the
$\cPqt$ tagging N-subjettiness requirement
on one selected jet (anti-tag), and determining the t tagging rate for the second jet (probe).
The anti-tag jet is required to satisfy $110<M_{\mathrm{SD}}<210\GeV$ and $\tau_{32}>0.69$.
This ``anti-tag and probe'' method yields a per-jet $\tmistag$  rate parameterized as a function of
jet momentum (which is more closely tied to the radiation within the jet than is the $\pt$) and is measured separately for events falling into each of the six $\cPqb$ tag
and $\abs{\Delta y}$ categories (Fig.~\ref{fig:mistag}).
The anti-tag requirement is designed to select a sample in data dominated by NTMJ events.
A small number of genuine \ttbar events survive this selection.
This contamination is removed by subtracting the distributions measured in \ttbar simulation from those measured in the anti-tag and probe selection in data.

Once the $\tmistag$ rate has been determined from the NTMJ control sample,
it is used to estimate the normalization and shape of NTMJ events passing the final event selection.
To do this, we use a ``single-tagged'' region that contains events with at least one $\cPqt$-tagged jet.
To avoid bias, we randomly select one of the two leading top quark jet candidates
and require that it pass the $\cPqt$ tagging selection described above.
If the randomly chosen jet is $\cPqt$ tagged, we include this event and weight it by the
appropriate $\tmistag$ rate based on the momentum of the jet opposite the tagged jet,
their rapidity difference, and the number of subjet $\cPqb$ tags, as shown in Fig.~\ref{fig:mistag}.

This singly-tagged control region without any requirements on the second jet has an overlap with the signal region, and is used to estimate the NTMJ background.
To remove the effects of double-counting, the \ttbar contribution is subtracted from the NTMJ estimate.
This is done by evaluating the $\tmistag$ weighting procedure described above on the simulated \ttbar events,
to find the contribution of \ttbar events that would enter the NTMJ background estimate when the method is applied to data. This contribution amounts to a \ttbar contamination of about 1--2\% of the NTMJ background estimate in the 0 $\cPqb$-tag event regions (about 6--10\% in the other regions), and is subtracted from the NTMJ background estimate.

As a final step in determining the shape of the NTMJ background estimate, we correct for the fact that the second jet,
having no $\ttagging$ applied, has different kinematics than jets in the signal region.
To mimic the kinematics of the signal region,
a ``mass-modified'' procedure is used, in which we randomly set the mass of this second jet according to a distribution of jet masses from simulated QCD multijet events, using the same window as used to select the signal region selection,
$110 < M_\mathrm{SD} < 210$\GeV.
This method is validated using simulated QCD multijet events.

\subsection{Systematic uncertainties}
\label{sec:systematics}

Several sources of systematic uncertainties are considered in this search.
Each of these is related to
 an experimental uncertainty introduced in the reconstruction of the event
or to a theoretical uncertainty affecting the simulation
of certain background or signal processes.
In particular, we quantify the effect of each of these uncertainties
on the measurement of the invariant mass of the reconstructed \ttbar system.
These uncertainties are taken into account in the maximum likelihood
fit to determine the total yield of SM processes, and in the statistical
interpretation of the data.
The complete list of systematic uncertainties is given below, and Table~\ref{tab:uncertainties}
lists the sources of uncertainty and the channels they affect.

The effect of the uncertainties in the theoretical SM cross sections
for $\ttbar$, $\Wjets$ and $\Zjets$ production
are obtained from the background fit described above, and
are
$ 8\%$ for $\ttbar$,
$ 6\%$ for $\Wjets$, and
20\% for $\Zjets$ production.
Small contributions to the event yields arise from single top quark and diboson production.
Their normalization is taken from
theory~\cite{Campbell:2011bn,Gehrmann:2014fva,Kidonakis:2013zqa,Aliev:2010zk,Kant:2014oha}
and assigned a 20\% uncertainty.
The effect due to missing higher-order corrections in the simulation
of \ttbar and $\Wjets$ production in the SM
is estimated by varying the renormalization
and factorization scales used in the simulation
up and down independently by a factor of 2. Additionally,
we account for uncertainties in the simulation of initial- and final-state radiation
on the reconstruction of the \ttbar system
by using \ttbar events simulated with different
$Q^2$ scales used for the parton shower generation and
evolution.
Simulated samples for both background and signal processes
are generated using PDFs from the NNPDF~3.0 set \cite{Ball:2014uwa}.
The corresponding systematic uncertainty
is determined according to
the procedure described in Ref.~\cite{Botje:2011sn}.
The uncertainty in the total integrated luminosity
at $\sqrt{s}=13\TeV$ is 2.7\%~\cite{CMS:LUM-15-001}.
The systematic uncertainty associated with
the yield of simulated pileup events
is evaluated by varying the inelastic pp cross section~\cite{CMS-PAS-FSQ-15-005}
by $\pm 5\%$ ($\sigma_{\text{inel}}=72.0 \pm 3.6~\text{mb}$).

The systematic uncertainties related to the muon identification and trigger efficiencies
are treated as uncorrelated,
and both are applied as functions of the muon $\pt$ and $\eta$~\cite{muonreco}.
The uncertainties are obtained by varying each corresponding data-to-simulation
SF by one standard deviation. Additional systematic uncertainties
of 1\% and 0.5\% are attributed to the identification and trigger efficiency SF measurements, respectively.
Similarly, the uncertainty in the electron identification efficiency
is applied as a function of the electron $\pt$ and $\eta$~\cite{electronreco}.
An uncertainty of 2\% is assigned to the efficiency
of the electron trigger selection, and
is determined from a complementary measurement
of the e+jets trigger efficiency in a dilepton ($\Pe\mu$) control region.

The uncertainties in the data-to-simulation corrections
for jet energy scale and jet energy resolution
are evaluated by varying these corrections within their uncertainties,
as functions of the jet $\pt$ and $\eta$.
Both systematic variations are also propagated
to the measurement of $\ptmiss$ and the jet mass.
A SF is applied to account for differing efficiencies and misidentification rates of the \btagging selection between data and simulation.
Uncertainties in the SFs
are measured as functions of the jet $\pt$ and treated as uncorrelated.
The data-to-simulation correction for the subjet $\btagging$ algorithm efficiency is
also included as an independent uncertainty and is evaluated
by varying the correction within its uncertainties, as a function of jet $\pt$ and $\eta$.
The data-to-simulation correction for the efficiency of
the $\ttagging$ selection for AK8 jets
is measured in situ in the statistical analysis.
This is done by leaving this parameter
unconstrained in the fit.
The $\tmistag$ efficiency
in the lepton+jets channel (dominated by quarks from $\Wjets$)
is measured directly in a control region dominated by $\Wjets$ production
with an uncertainty of 19\%.
The $\tmistag$ rate in the fully hadronic channel (dominated
by gluons from QCD interactions) is measured as described above, with a
momentum-dependent uncertainty ranging from 5 to 100\%. These
uncertainties are estimated by varying the anti-tag criterion for the
construction of the anti-tag and probe sample.
Systematic uncertainties due to the $\ttagging$ efficiency
and $\tmistag$ rate are treated as uncorrelated.

The systematic uncertainty associated with the ``mass-modified''
procedure, which is used to correct the kinematic bias in
the background estimation, is computed by taking half the
difference between the uncorrected and
``mass-modified'' background estimates. This affects the
shape and normalization of the $M_{\ttbar}$ distribution.
Simulated QCD multijet events are used in a closure test to verify
that the background estimation procedure accurately predicts
the double $\cPqt$-tagged $M_{\ttbar}$ distribution. An
additional systematic uncertainty is assigned to the NTMJ
background estimate based on small disagreements (up to 10\%)
observed in the closure test, in the shape of the kinematic threshold at low
values of $M_{\ttbar}$.

\begin{table}[htb]
\centering
\topcaption{Sources of uncertainty and the channels they affect. Uncorrelated uncertainties applied to a given channel are labeled with a $\odot$. Uncertainties that are correlated between channels are labeled with a $\oplus$. In this table, $\sigma$ denotes the uncertainty in the given prior value in the likelihood fit.}
\begin{tabular}{l c | c  c}
\multicolumn{2}{c}{{Uncertainty}} & \multicolumn{2}{c}{{Channel}}      \\
{Source} & {Prior uncertainty} & {Lepton+jets} & {Fully hadronic} \\
\hline
\ttbar cross section          & $\pm $8\%              & $\oplus$ & $\oplus$ \\
\Wjets cross section          & $\pm $6\%              & $\odot$  &          \\
\Zjets cross section          & $\pm $20\%             & $\odot$  &          \\
Single-top cross section      & $\pm $20\%             & $\odot$  &          \\
Diboson cross section         & $\pm $20\%             & $\odot$  &          \\
Integrated luminosity         & $\pm $2.7\%            & $\oplus$ & $\oplus$ \\
Pileup modeling               & $\pm 1 \sigma$         & $\oplus$ & $\oplus$ \\
Muon identification           & $\pm1\sigma(\pt,\eta)$ & $\odot$  &          \\
Muon trigger                  & $\pm1\sigma(\pt,\eta)$ & $\odot$  &          \\
Electron identification       & $\pm1\sigma(\pt,\eta)$ & $\odot$  &          \\
Electron trigger              & $\pm $2\%              & $\odot$  &          \\
Jet energy scale              & $\pm1\sigma(\pt,\eta)$ & $\oplus$ & $\oplus$ \\
Jet energy resolution         & $\pm1\sigma(\eta)$     & $\oplus$ & $\oplus$ \\
Jet b tagging efficiency      & $\pm1\sigma(\pt,\eta)$ & $\odot$  &          \\
Jet b mistag rate             & $\pm1\sigma(\pt,\eta)$ & $\odot$  &          \\
Subjet b tagging efficiency   & $\pm1\sigma(\pt,\eta)$ &          & $\odot$  \\
Jet t tagging efficiency      & unconstrained          & $\oplus$ & $\oplus$ \\
Lepton+jets channel t mistag rate    & $\pm19\%$              & $\odot$  &          \\
Fully hadronic channel t mistag rate    & $\pm1\sigma(p)$        &          & $\odot$  \\
PDFs                 & $\pm 1 \sigma$         & $\oplus$ & $\oplus$ \\
\ttbar matrix element scale   & $\pm 1 \sigma$         & $\oplus$ & $\oplus$ \\
\ttbar parton shower scale    & $\pm 1 \sigma$         & $\oplus$ & $\oplus$ \\
\Wjets matrix element scale   & $\pm 1 \sigma$         & $\odot$  &          \\
NTMJ background kinematics           & $\pm 1 \sigma$         &          & $\odot$  \\
NTMJ background closure test             & $\pm 1 \sigma$         &          & $\odot$  \\
\end{tabular}
\label{tab:uncertainties}

\end{table}

\subsection{Fitting procedure}
\label{sec:fitting}

To improve the flexibility of the background model, we
estimate the central values and uncertainties in several parameters through
a maximum likelihood fit to data using the top quark pair invariant mass distribution, as follows.
The normalizations for the background estimates using simulated events are left unconstrained in the
fit. The data-to-simulation SF for the
$\ttagging$ efficiency is also unconstrained and extracted from the
fit.  The SF for the subjet b tagging
efficiency as well as the yield of events from the NTMJ background estimation method, having both $\pt$ and $\eta$ dependence, are
allowed to vary within uncertainties, with their final values
estimated by the fit.
The NTMJ background is constrained using the procedure outlined
in Section \ref{sec:bkg_allhad}.
All other systematic uncertainties
are included as nuisance parameters in the fit, and are allowed to vary within their corresponding rate and shape uncertainties, as described above, using log-normal prior distributions.
The best fit values obtained from this maximum likelihood evaluation
are used to correct the distributions of background and signal processes.

A Bayesian statistical method~\cite{bayesbook, theta} is used to extract
the upper limits at 95\% confidence level (CL) on the product of the cross section
and branching fraction, i.e.
$\sigma(\Pp\Pp\to X) \, \mathcal{B}(X\to \ttbar)$,
for heavy resonances decaying to a \ttbar pair.
In order to maximize the expected sensitivity of the search,
twelve exclusive categories are employed simultaneously in the
statistical analysis, as described above.
For each category, the observable used in the limit-setting procedure
is $\mttbar$.  A template-based shape analysis is performed using the Theta software
package \cite{theta} for these $M_{\ttbar}$ distributions.
The systematic uncertainties listed in Table \ref{tab:uncertainties}
are introduced as individual nuisance parameters in the limit calculation.
For the signal cross section parameter, we use a uniform prior distribution.
The uncertainty in the data-to-simulation correction
for $\ttagging$ efficiency is left unconstrained,
whereas each of the other nuisance parameters corresponding to a
systematic uncertainty is modeled with
a log normal prior distribution.
The uncertainty due to the finite size of the simulated samples
is introduced in the statistical analysis according
to the ``Barlow--Beeston lite'' method~\cite{barlow_beeston}.
The impact of the statistical uncertainty in the simulated samples is
limited by rebinning each $\mttbar$ distribution to ensure that the statistical uncertainty
associated with the expected background is less than 30\% in each bin.

\begin{table}[h!tb]
\centering
\topcaption{\label{tab:event_yields__SR}
   Numbers of events in the signal region for the lepton+jets analysis.
   The expected yields for SM backgrounds are obtained from the
   maximum likelihood fit to the data described in Section \ref{sec:fitting}.
   The uncertainties reported in the total expected background include
   the statistical uncertainties in the simulation and
   all the posterior systematic uncertainties.  For the W+jets background, LF (HF) indicates contributions from W bosons produced in association with light-flavor (heavy-flavor) jets.
  }
\begin{tabular}{lccc}
                 & \multicolumn{3}{c}{{$\mu$+jets signal region}} \\
{Process} & { 1 t tag } & { 0 t tags, 1 b tag } & { 0 t tags, 0 b tags }  \\
\hline

$\ttbar$ & $218 \pm 28$ &        $7602 \pm 826$ &        $1965 \pm 229$ \\
$\Wjets$ (LF) &  $27 \pm  4$ &         $547 \pm  54$ &        $4675 \pm 377$ \\
$\Wjets$ (HF) &   $4 \pm  1$ &         $333 \pm  30$ &         $780 \pm  65$ \\
Other &   $9 \pm  2$ &         $682 \pm 111$ &         $635 \pm  85$ \\
\hline
Total background & $258 \pm 29$ &        $9164 \pm 856$ &        $8055 \pm 541$ \\
Data & 252        &        9230         &        7966         \\

\end{tabular}
\vspace{1.0 cm}

\begin{tabular}{lccc}
                 & \multicolumn{3}{c}{{e+jets signal region}} \\
{Process} & { 1 t tag } & { 0 t tags, 1 b tag } & { 0 t tags, 0 b tags }  \\
\hline

$\ttbar$ & $119 \pm 15$ &        $1016 \pm 124$ &          $248 \pm 32$ \\
$\Wjets$ (LF) &  $13 \pm  2$ &          $97 \pm  10$ &          $684 \pm 58$ \\
$\Wjets$ (HF) &   $2 \pm  1$ &          $44 \pm   4$ &           $84 \pm  8$ \\
Other &   $4 \pm  1$ &         $103 \pm  18$ &           $74 \pm 10$ \\
\hline		
Total background & $138 \pm 16$ &        $1260 \pm 129$ &         $1090 \pm 78$ \\
Data & 142        &        1217         &         1005        \\

\end{tabular}

\end{table}

\begin{table}[htb!]
\centering
\topcaption{Number of events in the signal region for the fully hadronic analysis.
   The expected yields for SM backgrounds are obtained from the
   maximum likelihood fit to data described in the text.
   The uncertainties reported for the total expected background include
   the statistical uncertainties on the simulation and
   all the posterior systematic uncertainties.}
\begin{tabular}{lccc}
                 & \multicolumn{3}{c}{{$\abs{\Delta y} > 1.0$ signal region}} \\
{Process} & { 0 b tags } & { 1 b tag } & { 2 b tags }  \\
\hline
SM \ttbar          & $   34 \pm 4.3  $ & $  62 \pm 5.8  $ & $  28 \pm 3.8 $  \\
NTMJ               & $  787 \pm 6.2  $ & $ 215 \pm 4.7  $ & $  15 \pm 1.9  $  \\
\hline

Total background   & $  821 \pm 7.5  $ & $  278 \pm 7.4  $ & $  43 \pm 4.2  $ \\
Data      & $  830   $      & $   264        $ &   $ 46   $      \\

\end{tabular}
\vspace{1.0 cm}

\begin{tabular}{lccc}
                 & \multicolumn{3}{c}{{$\abs{\Delta y} < 1.0$ signal region}} \\
{Process} & { 0 b tags } & { 1 b tag } & { 2 b tags }  \\
\hline
SM \ttbar          & $  66 \pm 7.1  $ & $  121 \pm 10  $ & $  60 \pm 7.0  $  \\
NTMJ               & $  817 \pm 8.0  $ & $  248 \pm 7.0  $ & $  19 \pm 1.7  $  \\
\hline

Total background   & $  882 \pm 11  $ & $  369 \pm 12  $ & $  79 \pm 7.3  $ \\
Data      & $  925   $      & $   387        $ & $   94$        \\
\end{tabular}
\label{tab:event_yields__allhad}

\end{table}

\begin{figure}[htbp]
 \centering
 \includegraphics[width=0.45\textwidth]{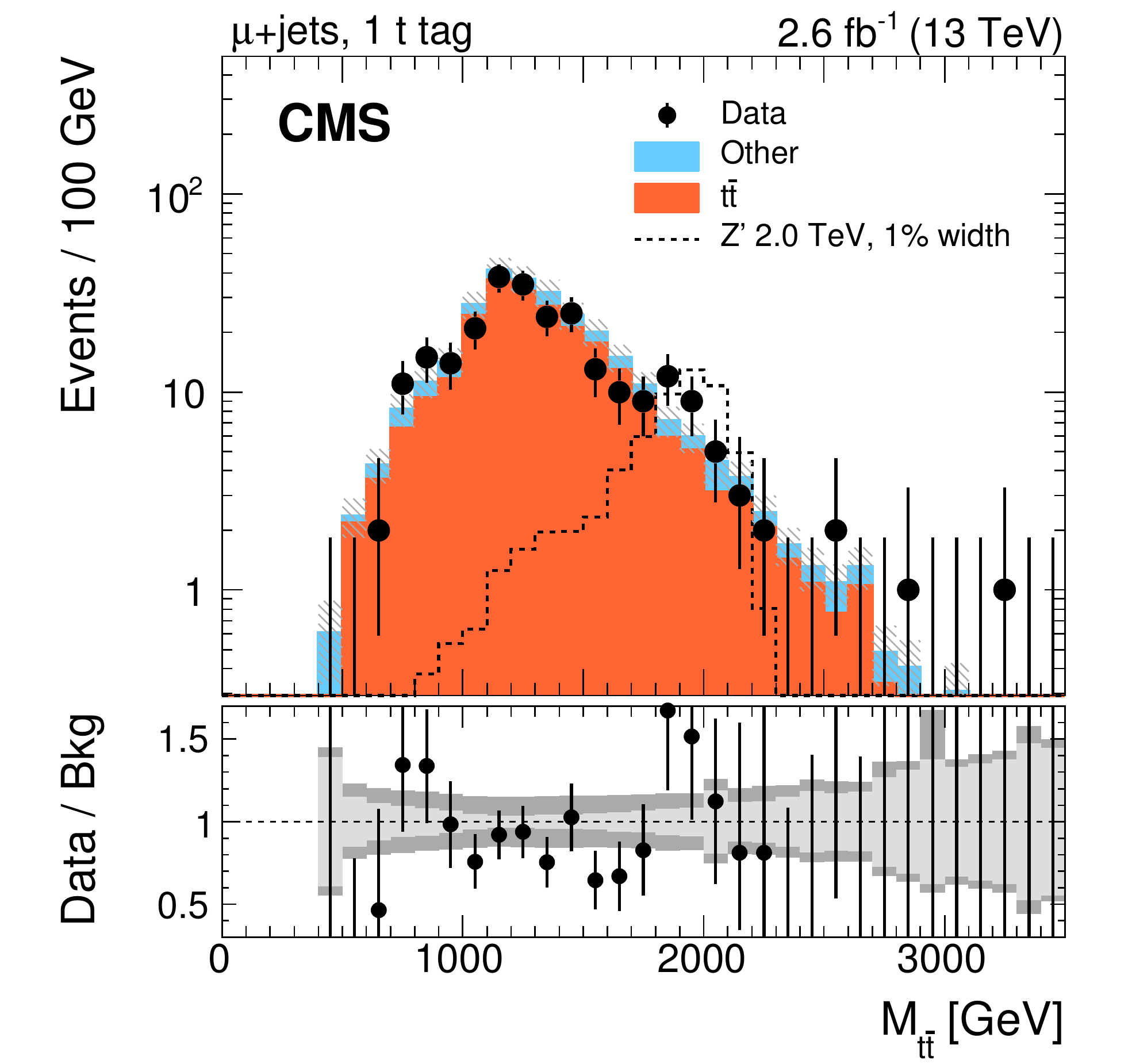}
 \includegraphics[width=0.45\textwidth]{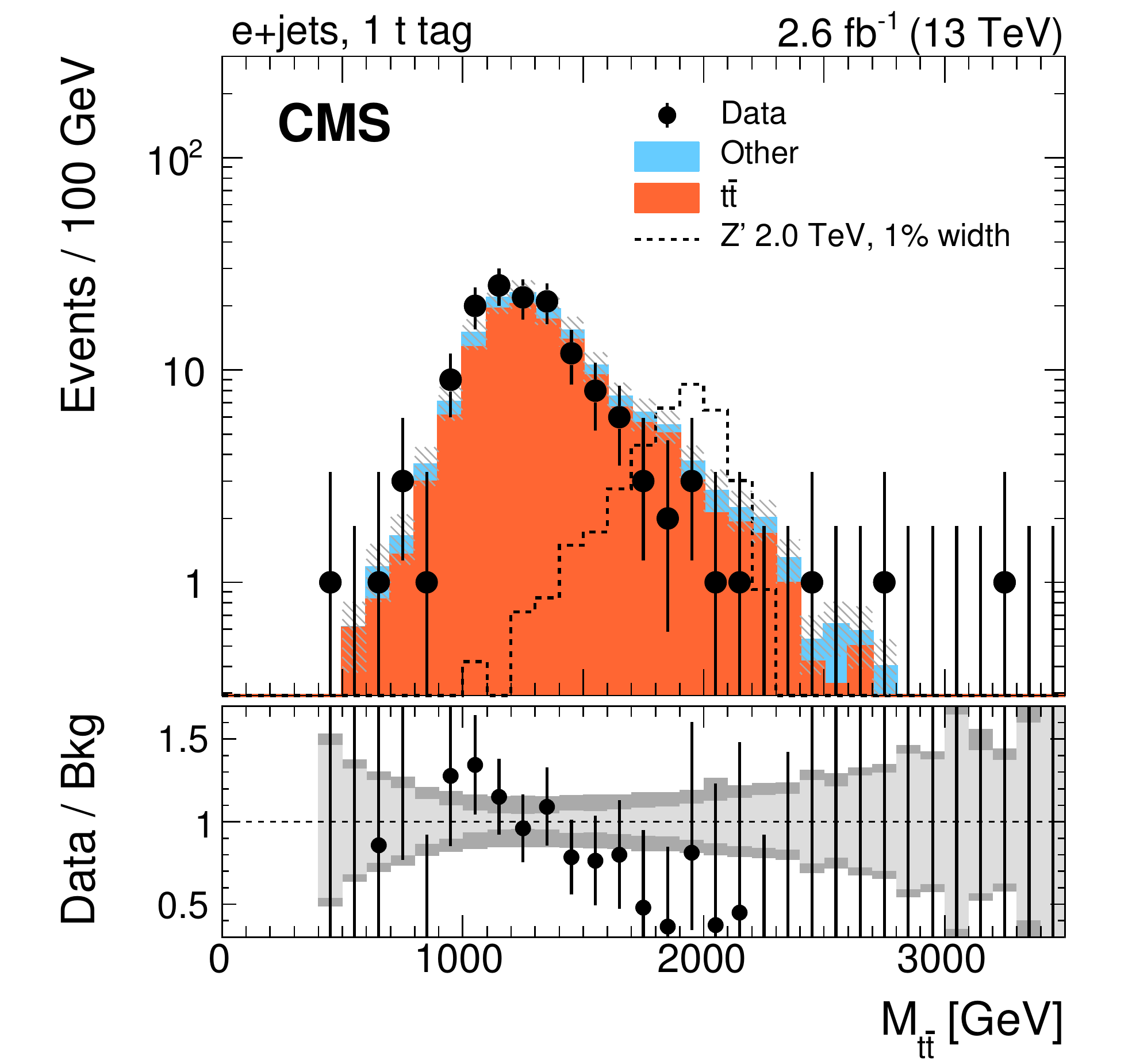}
 \includegraphics[width=0.45\textwidth]{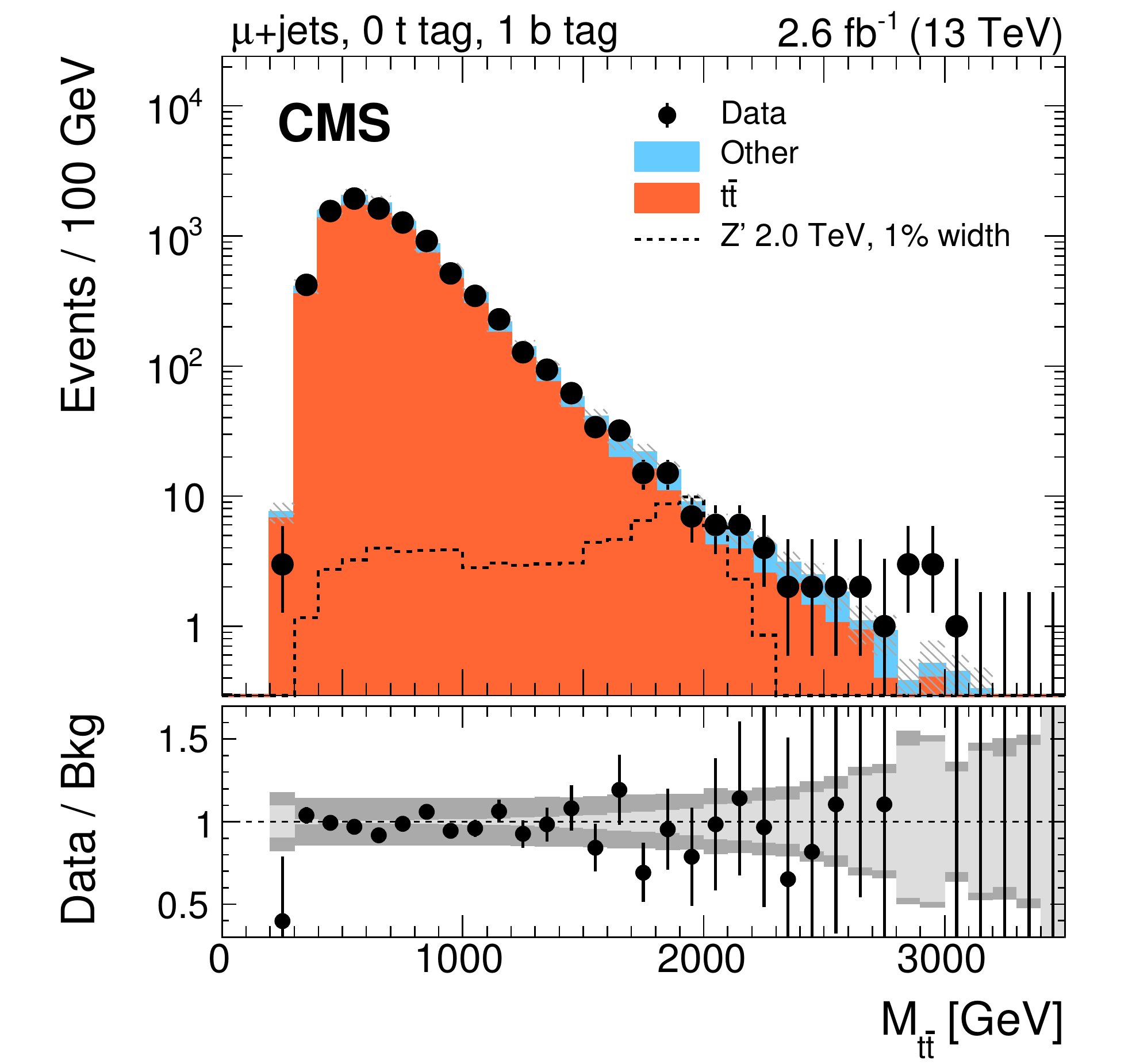}
 \includegraphics[width=0.45\textwidth]{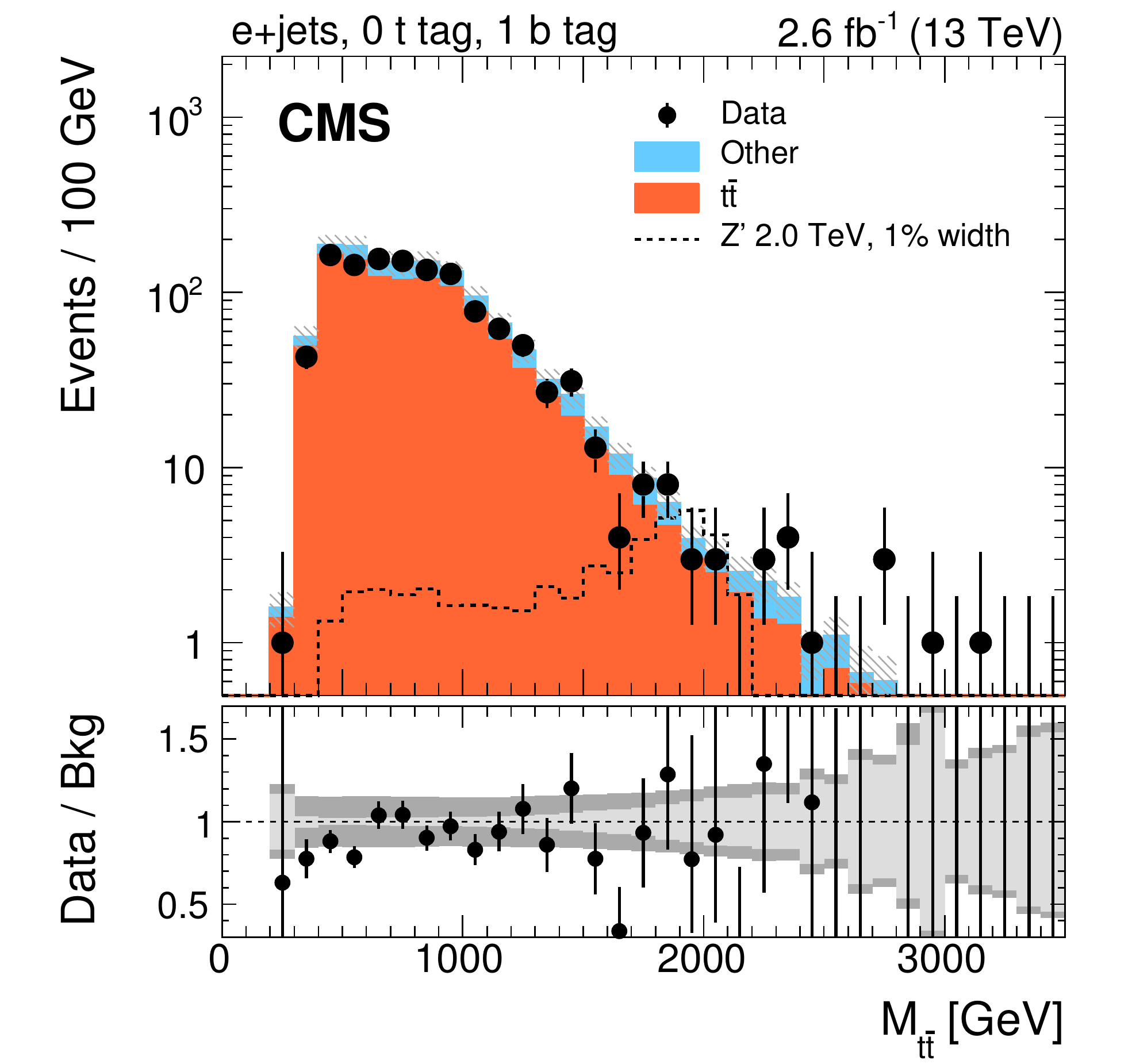}
 \includegraphics[width=0.45\textwidth]{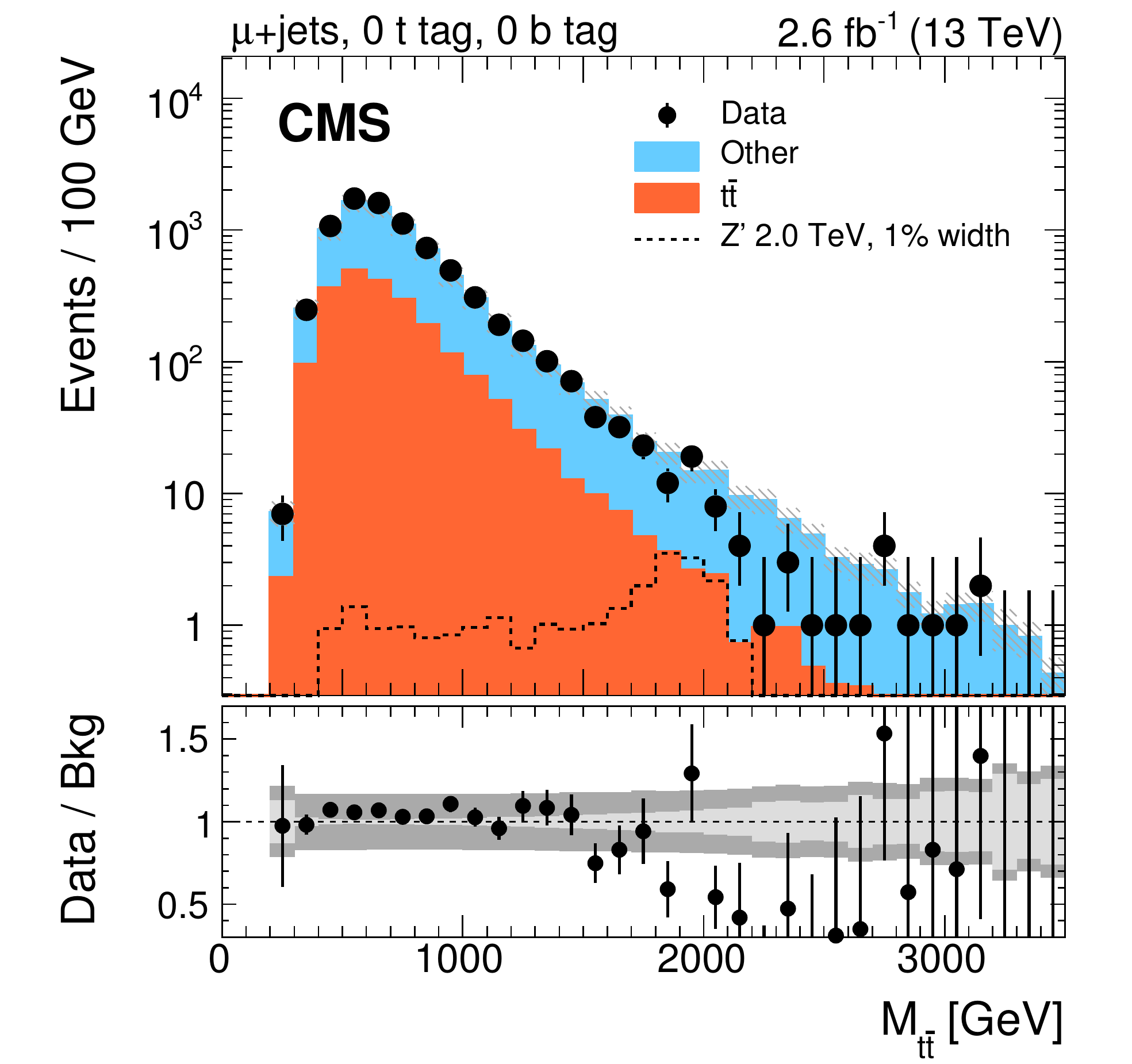}
 \includegraphics[width=0.45\textwidth]{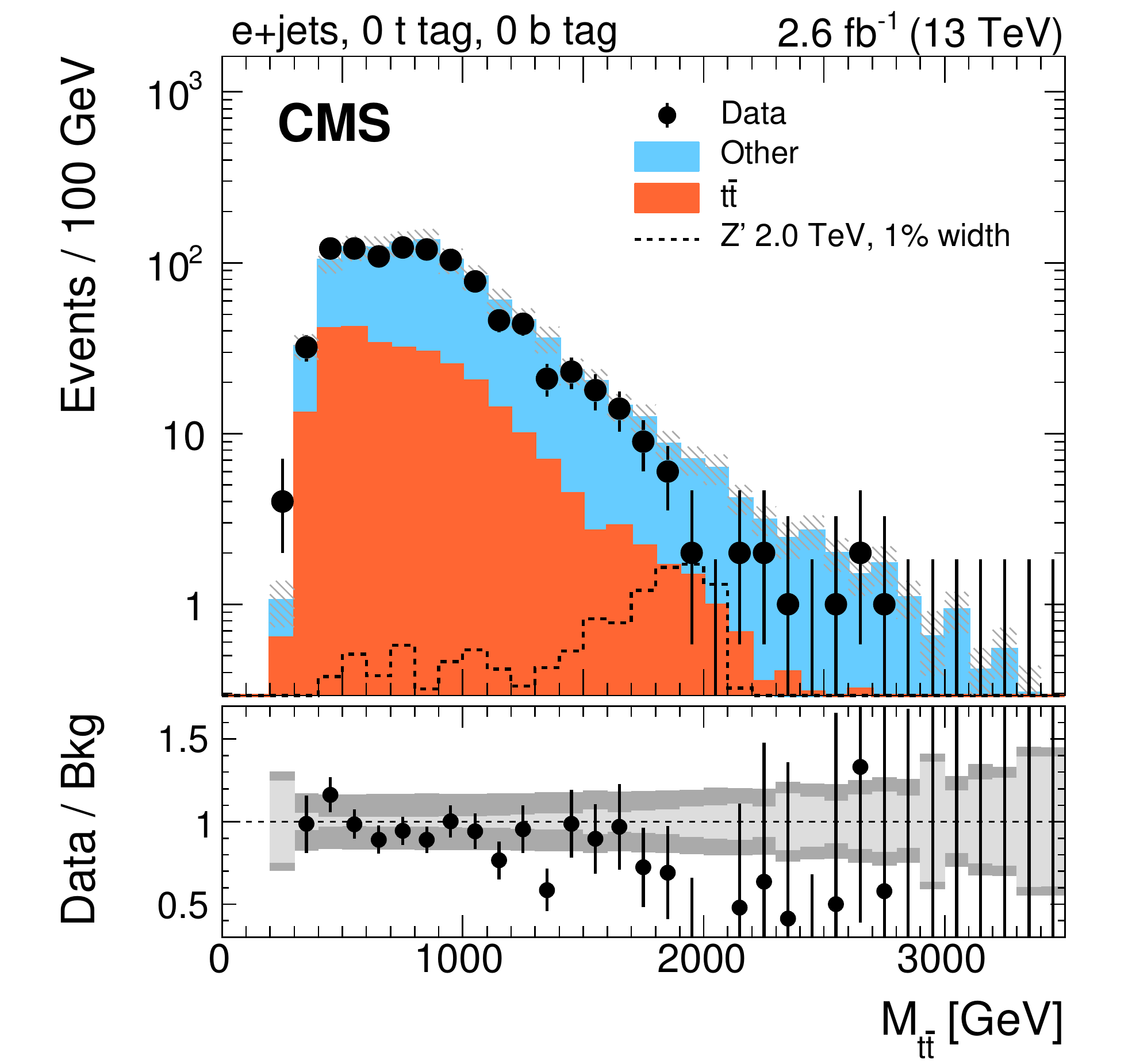}
 \caption{Distributions in $\mttbar$
          for data and expected background, for events passing
          the signal selection of the lepton+jets analysis ($\chi^{2}<30$) after the maximum likelihood fit.
          Distributions are shown for the muon (left) and electron (right) channel.
          For each lepton flavor, events are split into three exclusive categories
          (from uppermost to lowest):
          $\It$,
          $\OtIb$, and
          $\OtOb$.
          The signal templates are normalized to a cross section of 1\unit{pb}.
          The uncertainties associated with the background expectation include
          the statistical and all post-fit systematic uncertainties.
          The lower panel in each figure shows the ratio of data to predicted SM background, with
          the statistical (light gray) and
          total (dark gray) uncertainties shown separately.}
 \label{fig:mttbar__SR__postfit}
\end{figure}

\begin{figure}
\centering
\includegraphics[width=0.45\linewidth]{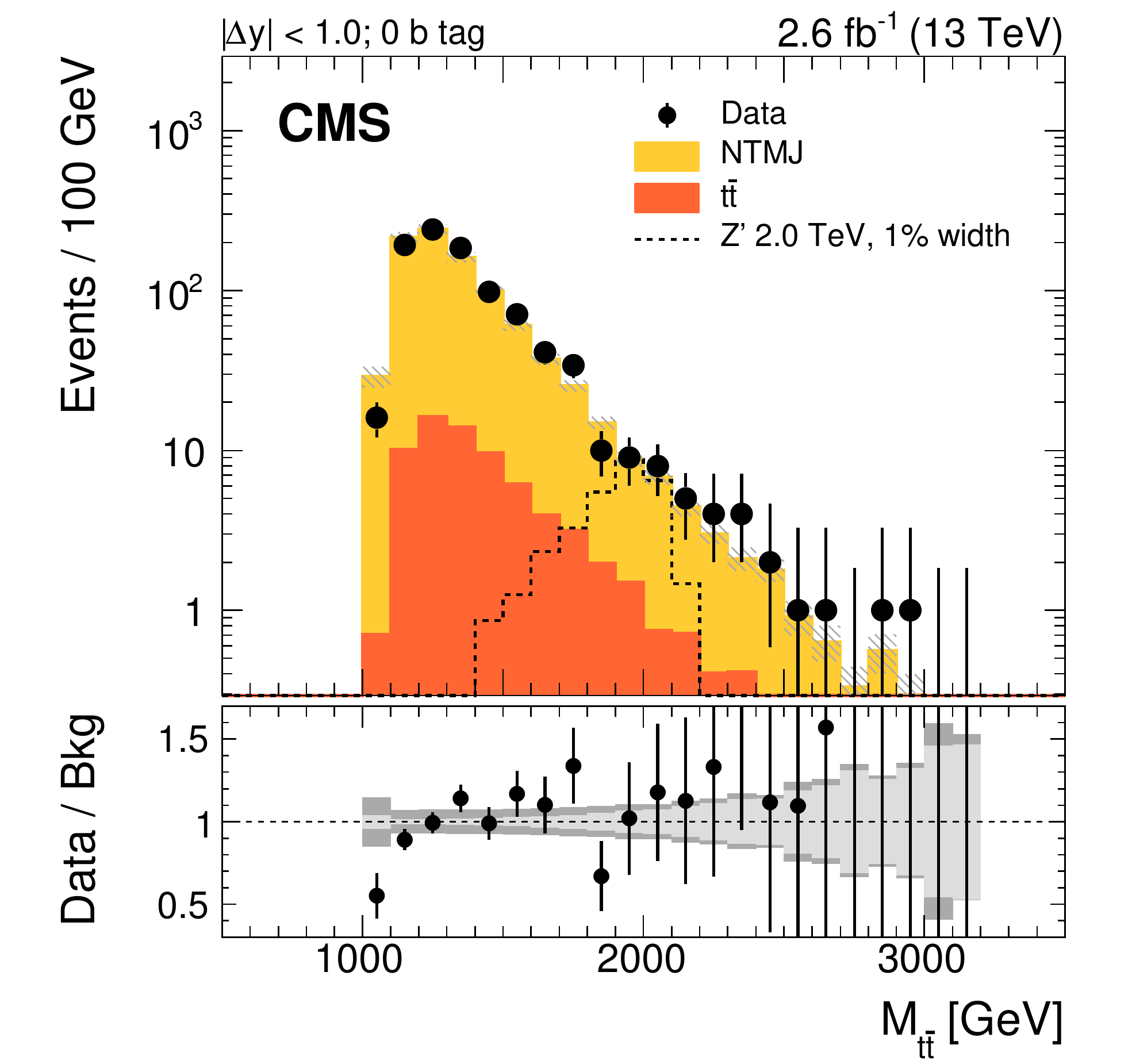}
\includegraphics[width=0.45\linewidth]{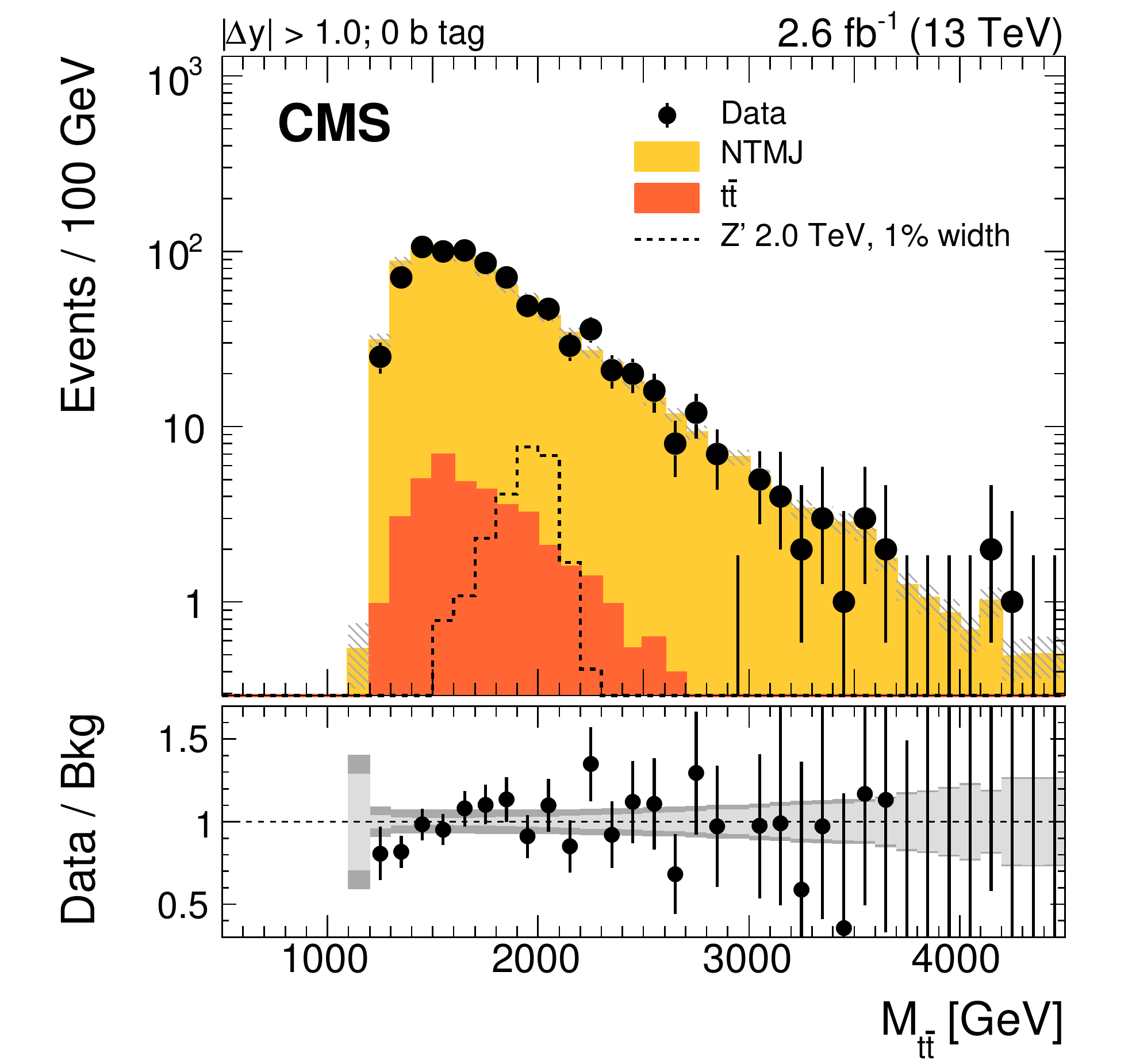}
\includegraphics[width=0.45\linewidth]{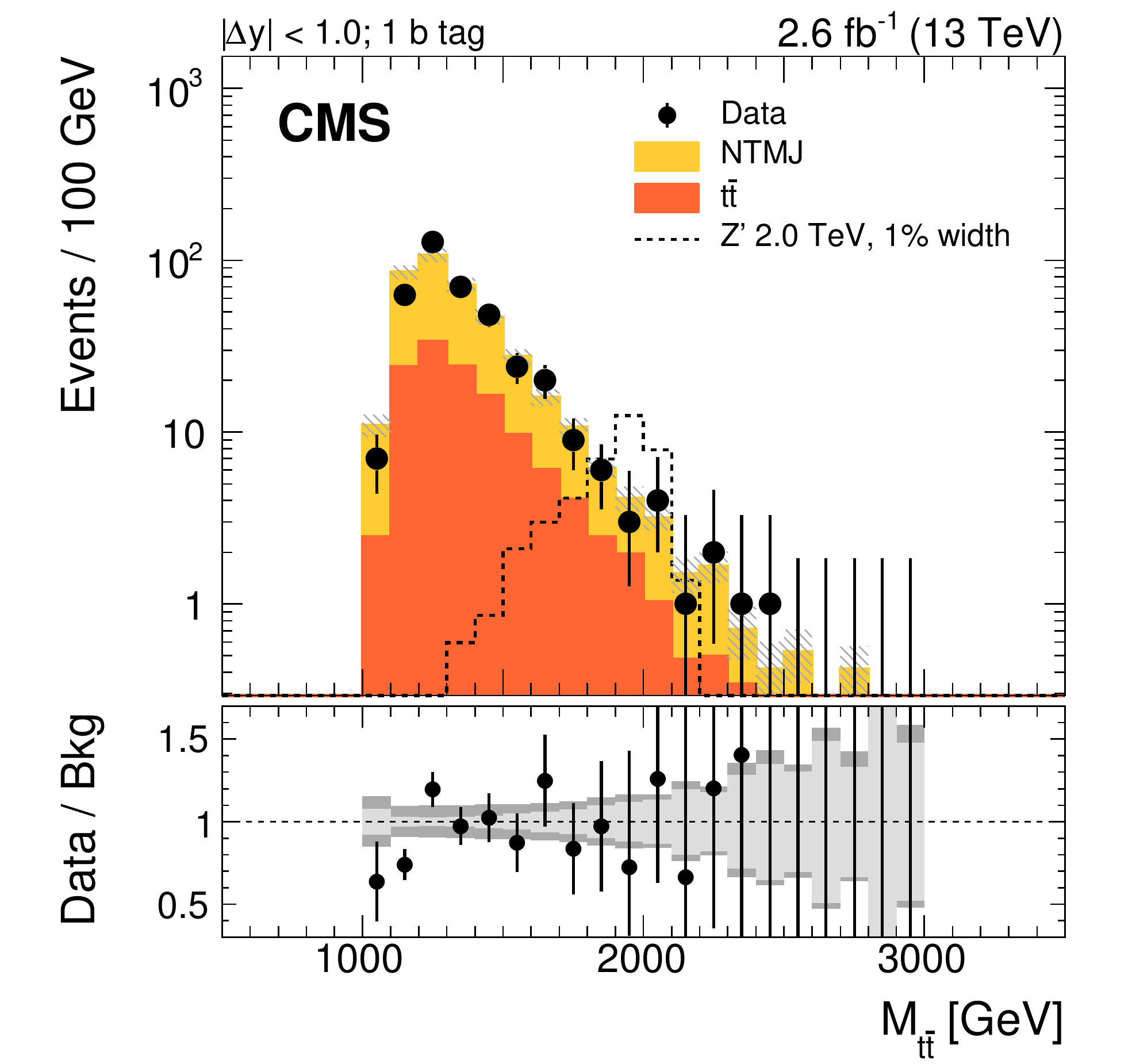}
\includegraphics[width=0.45\linewidth]{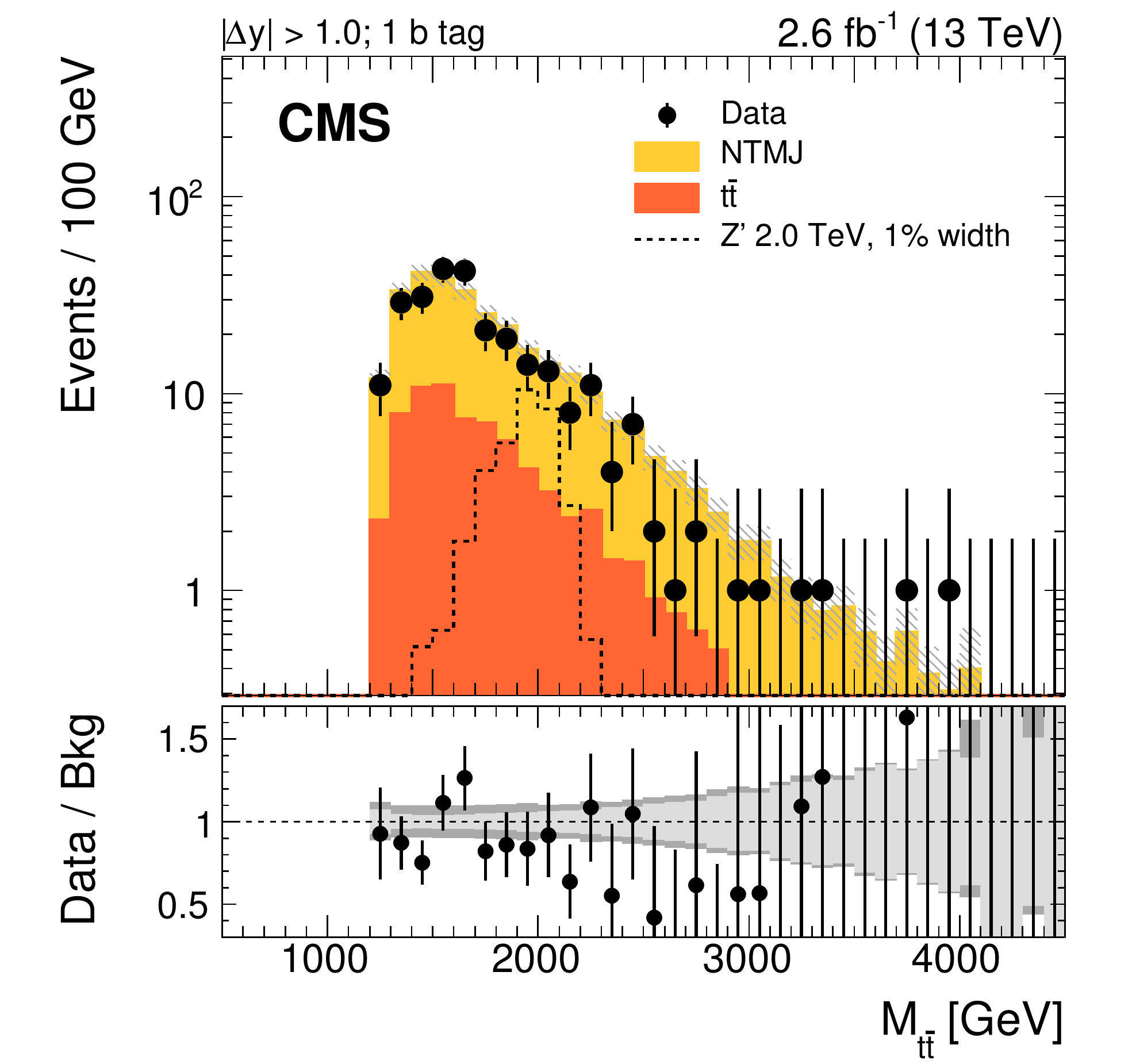}
\includegraphics[width=0.45\linewidth]{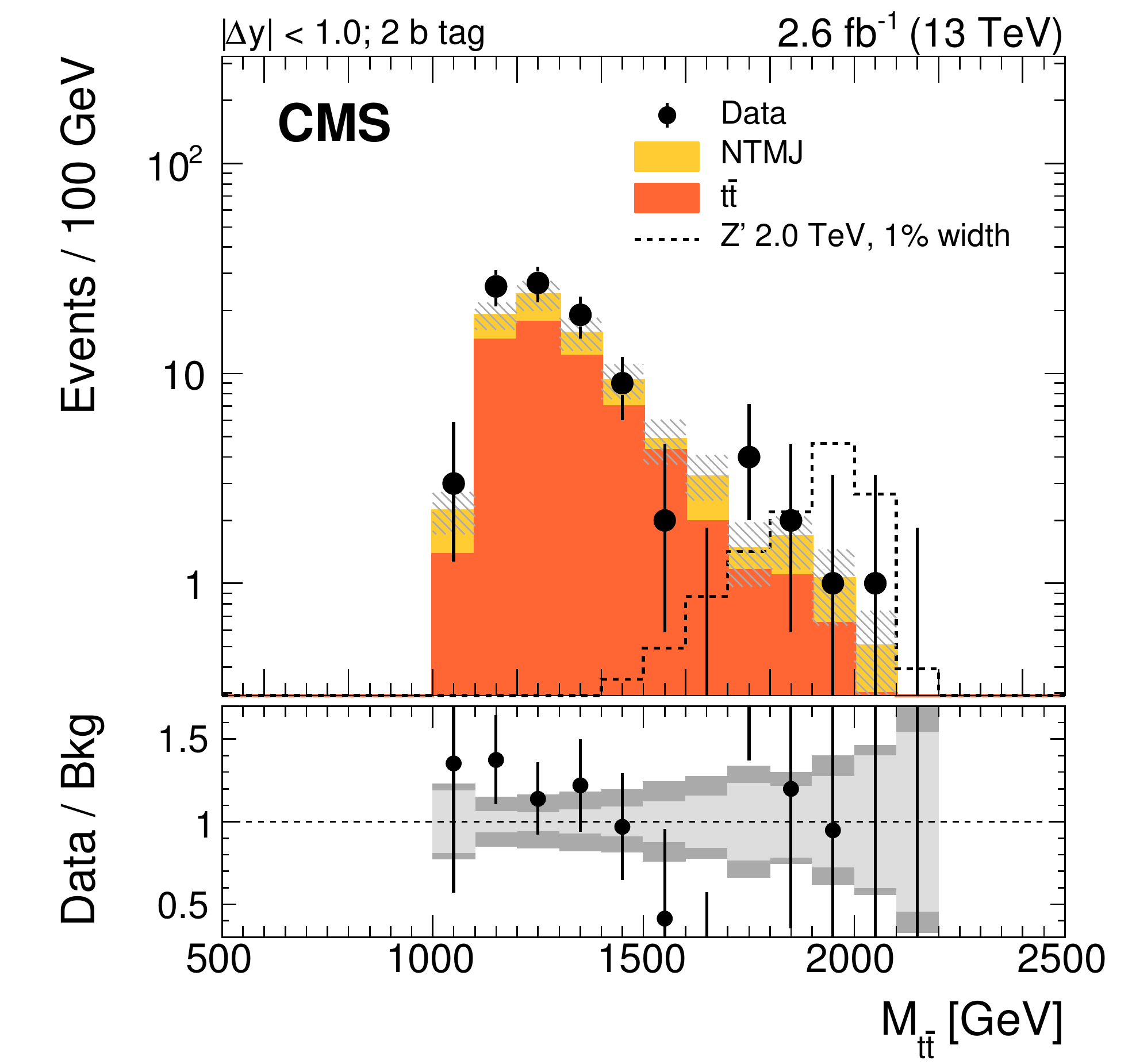}
\includegraphics[width=0.45\linewidth]{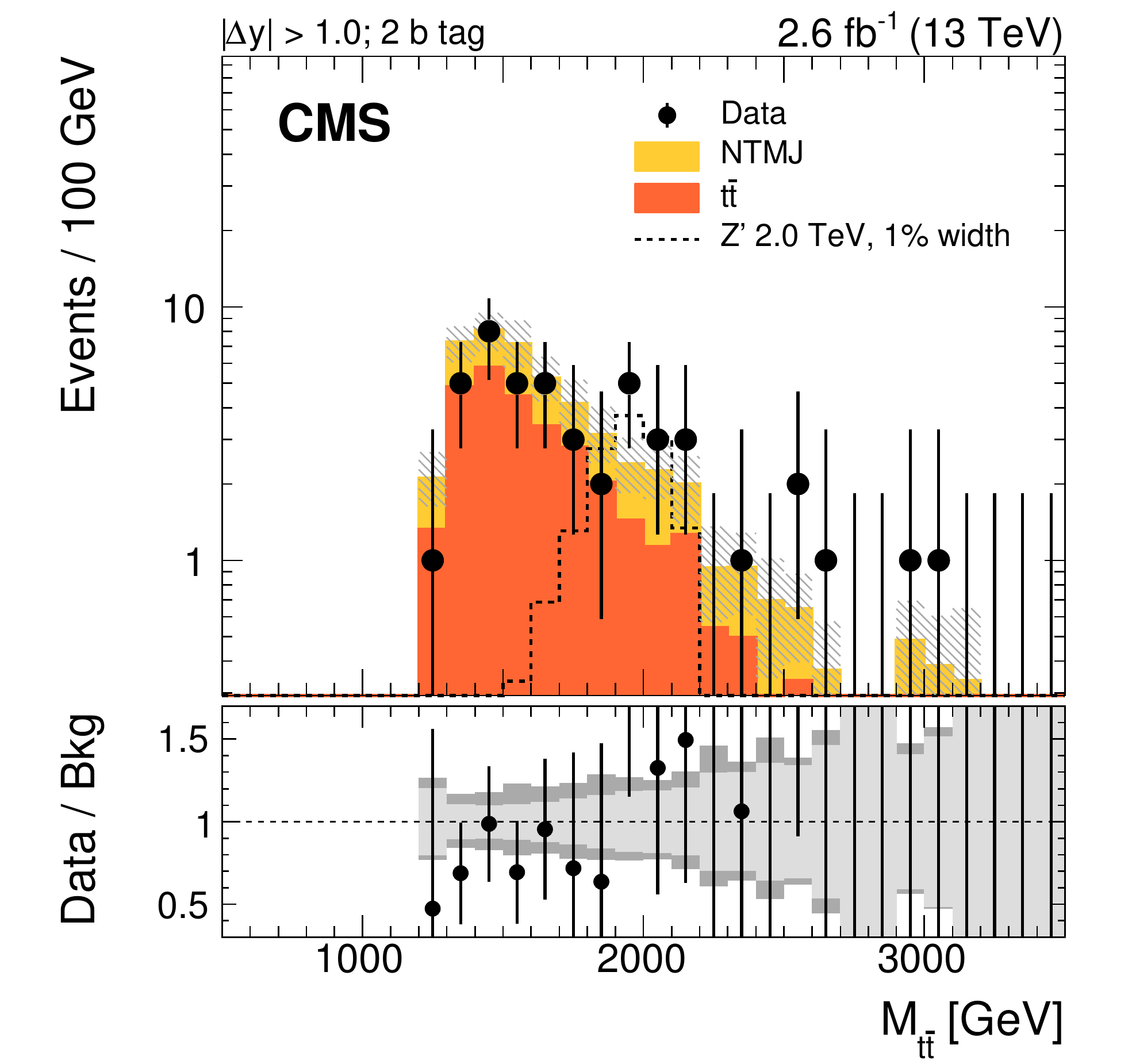}
\caption{
         Distributions in $\mttbar$
          for data and expected background, for events passing
          the signal selection of the fully hadronic analysis after the maximum likelihood fit.
          Distributions are shown for the regions with $\abs{\Delta y} <
          1.0$ (left) and $\abs{\Delta y} > 1.0$ (right), for 0, 1, or 2
          subjet b tags (from uppermost to lowest).
          The signal templates are normalized to a cross section of 1\unit{pb}.
          The uncertainties associated with the background expectation include
          the statistical and all post-fit systematic uncertainties.
          The lower panel in each figure shows the ratio of data to predicted SM background, with the statistical (light gray) and total (dark gray)
	  uncertainties shown separately.
}
\label{fig:final_mtt_allhad}

\end{figure}

\section{Results}
\label{sec:results}

The number of events observed in data and
expected from SM processes after the background fit
are given in Tables \ref{tab:event_yields__SR}
and \ref{tab:event_yields__allhad} for the six
categories in the signal region of the lepton+jets
and fully hadronic channels, respectively.
The invariant mass distribution of the reconstructed \ttbar pair
is shown in~\figref{fig:mttbar__SR__postfit} (\figref{fig:final_mtt_allhad})
for data and the expected SM backgrounds in the lepton+jets (fully hadronic) signal-region categories
after the background fit.
Good agreement between data and background prediction is observed within the estimated systematic uncertainties.
The modeling of the data in background-enriched samples is verified using kinematic distributions for leptons, jets, and the reconstructed leptonically and hadronically decaying top quarks in each of the individual categories considered in the analysis.
The small differences are covered by the systematic uncertainties.  For the lepton+jets channel, some discrepancies are observed at large $\mttbar$ in the distributions in categories where the W+jets background dominates.  These discrepancies are related to missing higher-order corrections in the simulated events, and have little impact on the results, as these categories are less sensitive than those dominated by \ttbar.  Dedicated cross checks have confirmed that the localized discrepancies visible in Figs.~\ref{fig:mttbar__SR__postfit} and \ref{fig:final_mtt_allhad} may be attributed to statistical fluctuations.  The sensitivity of this analysis is driven by the 1 t tag categories in the lepton+jets channel, and the 2 b tag categories in the fully hadronic channel, which have the highest signal-to-background ratios.

We proceed to set exclusion limits on different benchmark models for \ttbar resonances.
Four extensions to the SM are considered in the statistical analysis:
a $\PZpr$ boson decaying exclusively to \ttbar with a relative decay width ($\Gamma/M$) of 1\%, 10\%, or 30\%,
and a KK gluon resonance in the RS model.
The cross sections for $\PZpr$ production are taken from NLO
order calculations~\cite{Bonciani:2015hgv}.
The leading order (LO) predictions for the KK gluon cross sections
are multiplied by a factor of 1.3
to account for higher-order corrections~\cite{Gao:2010bb}.

\begin{figure}[htb]
\centering
\begin{tabular}{cc}
\includegraphics[width=0.45\linewidth]{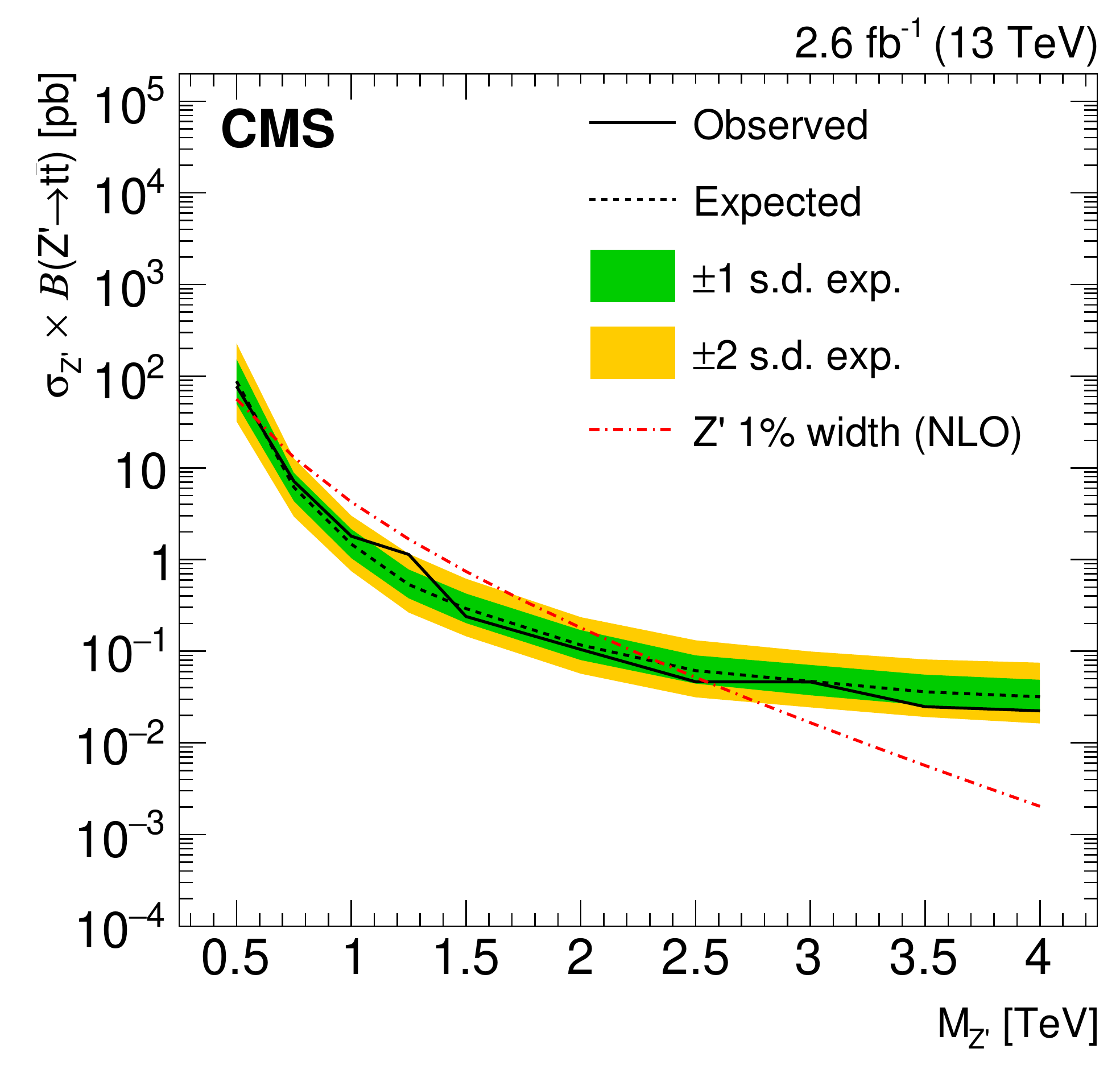}
\includegraphics[width=0.45\linewidth]{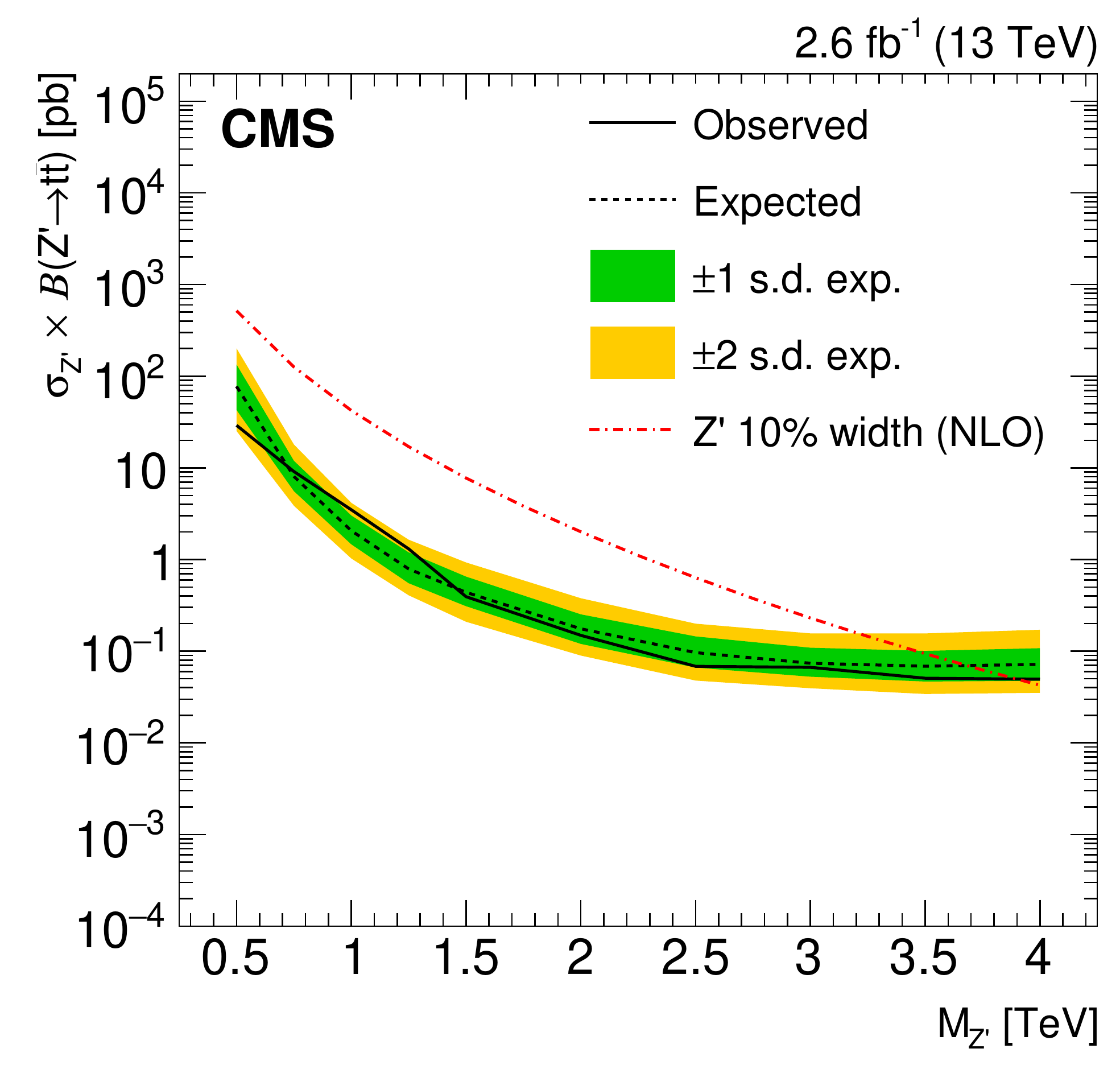} \\
\includegraphics[width=0.45\linewidth]{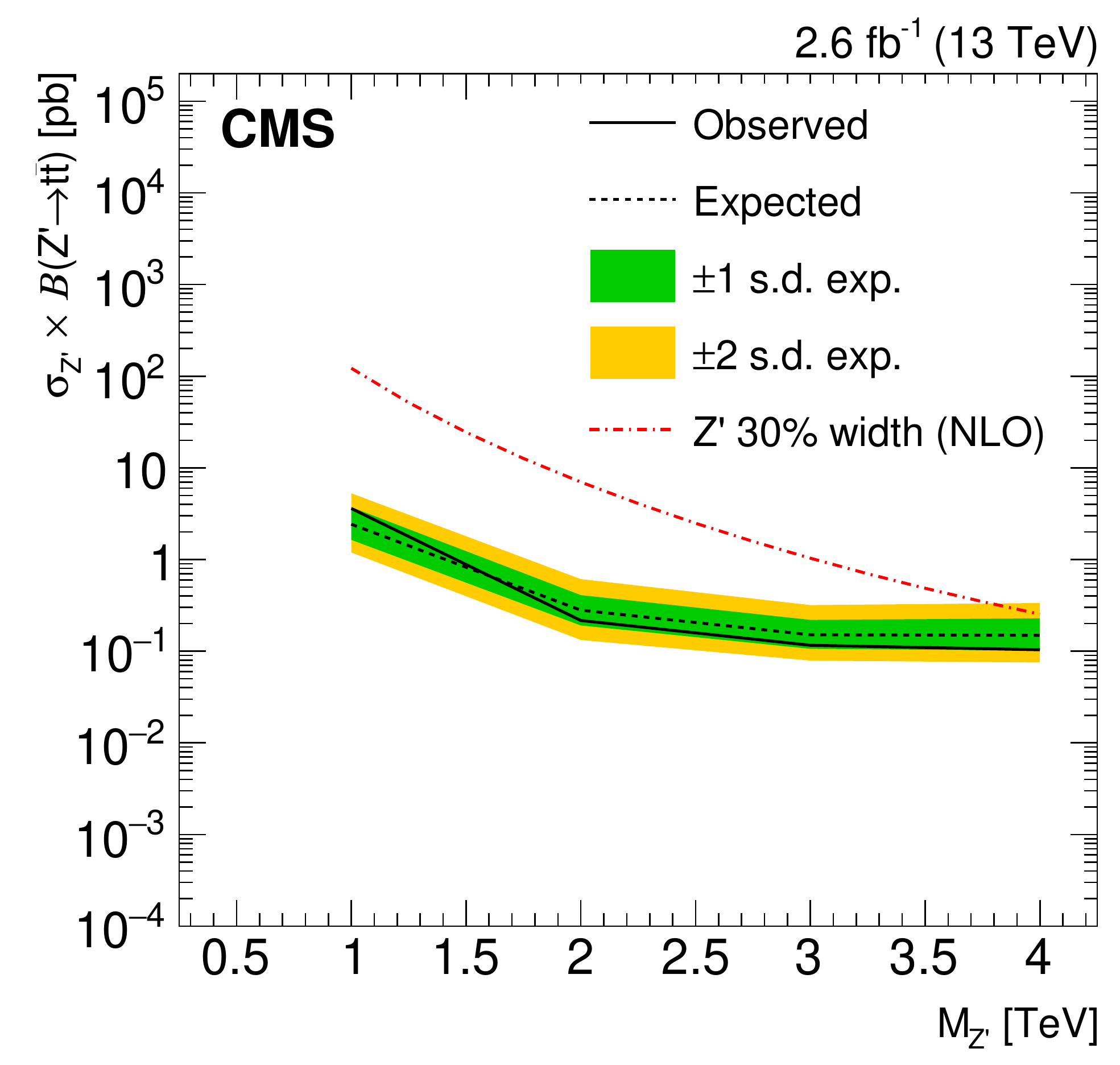}
\includegraphics[width=0.45\linewidth]{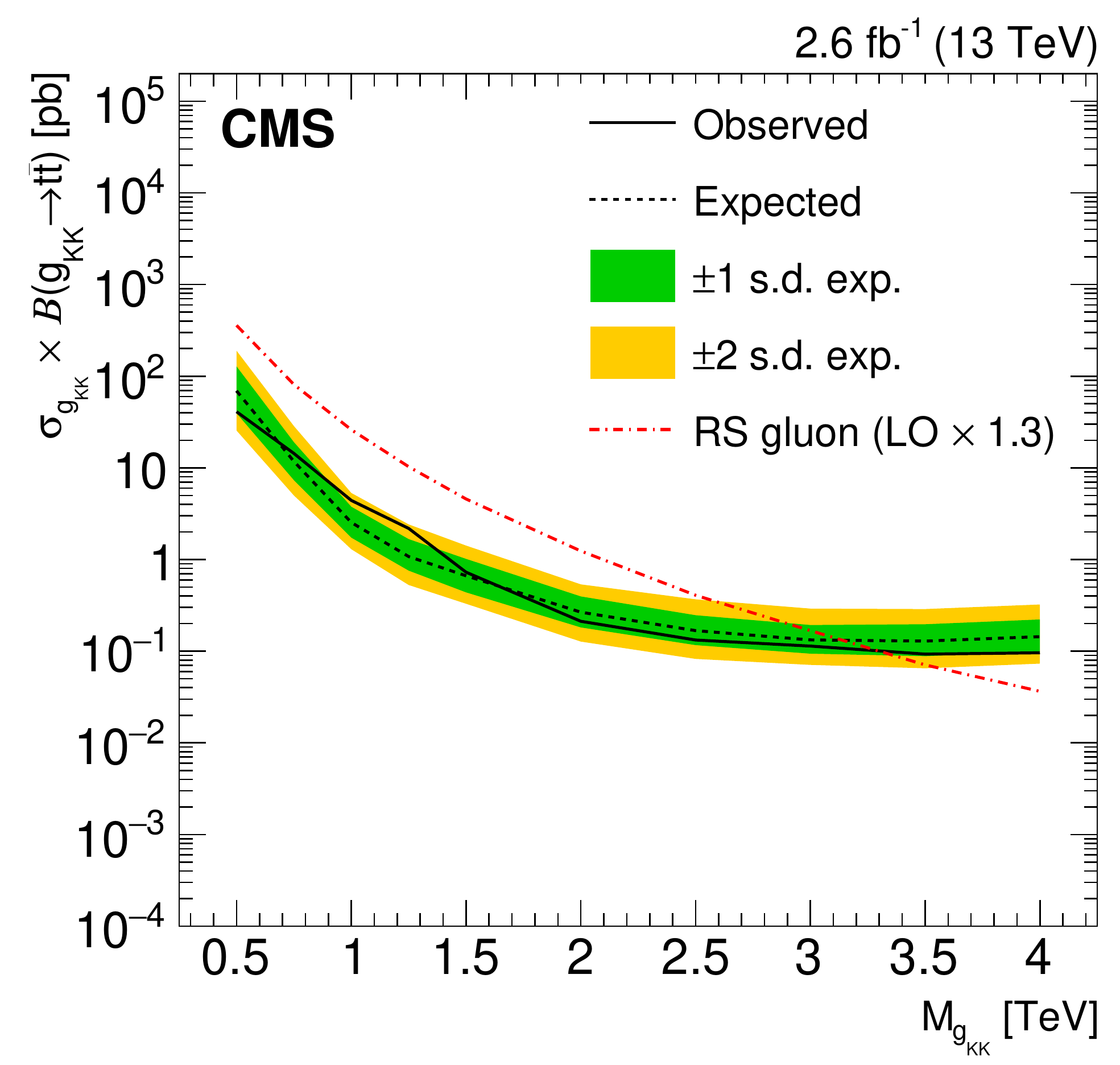}  \\
\end{tabular}
\caption{Observed and expected upper limits at 95\% CL on the product of the production cross section and branching fractions for the full combination of the analysis results, shown as function of the resonance mass. Limits are set using four extensions to the SM :
          (upper left) a $\PZpr$ boson with $\Gamma/M$ of  1\%,
          (upper right) a $\PZpr$ boson with $\Gamma/M$ of 10\%,
          (lower left) a $\PZpr$ boson with $\Gamma/M$ of 30\% and
          (lower right) a KK excitation of a gluon in the RS model.  The corresponding theoretical prediction as a function of the resonance mass is shown as a dot-dashed curve.}
\label{combo}

\end{figure}

\begin{figure}[htb]
\centering
\includegraphics[width=0.90\textwidth]{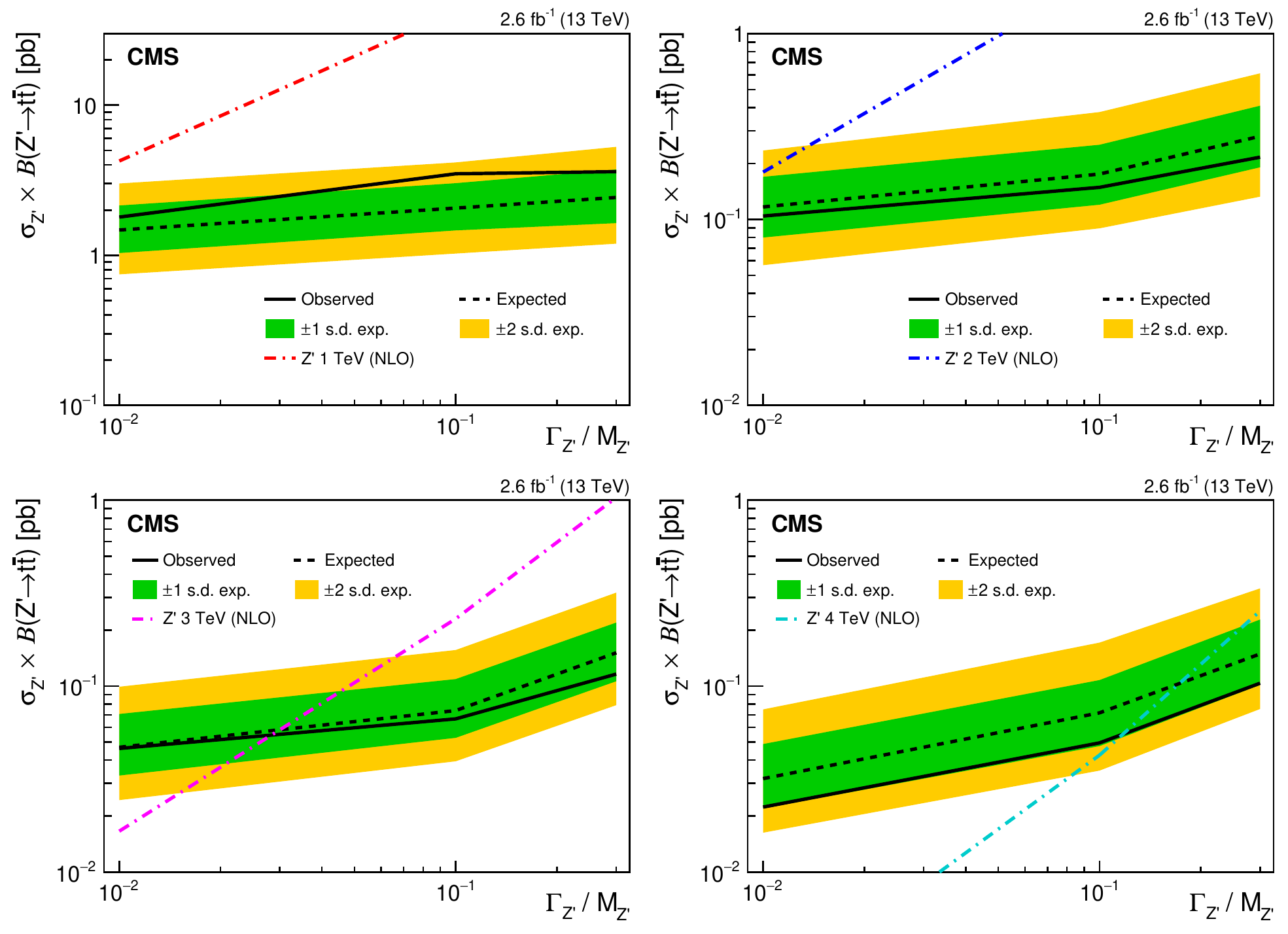}
\caption{Expected and observed limits presented as a function of width, for $M_{\PZpr} = 1,$ 2, 3, 4\TeV.  The corresponding theoretical prediction as a function of width is shown as a dot-dashed curve in each case.}
\label{width}
\end{figure}

Limits are extracted on
the cross sections for the various signal hypotheses using the distributions in Figs.~\ref{fig:mttbar__SR__postfit} and~\ref{fig:final_mtt_allhad}.  By varying the nuisance parameters within their prior distribution
functions, pseudo-experiments are performed to estimate the 68\% and
95\% CL (1 and 2 standard deviations) expected limits in the median results.
The combined results, including observed limits on the
resonant production cross sections, are shown in Fig.~\ref{combo}, and
tabulated in Tables \ref{compare_results}--\ref{tab_kkg}.
The combination of the lepton+jets and fully hadronic channels significantly improves the
exclusion limits relative to previous results for all models, except for those using a
width of 1\%.  Starting from the lower mass exclusion limit of 0.5\TeV, masses are excluded
up to 4\TeV for the 30\% width \PZpr samples, up to 3.9\TeV for the 10\% width \PZpr, and up to 3.3\TeV for the RS KK gluon hypotheses, at the 95\% CL.
These limits are close to the point where the parton luminosity at low
\ttbar mass dominates the mass distribution by enhancing the off-shell contribution
and reducing the resonant contribution, modifying the behavior of the signal model from
resonant-like to nonresonant-like. Because of this,
a different analysis strategy should be considered in future searches, in order to be sensitive to such non-resonant
production at large $\mttbar$.
Table \ref{compare_results} shows the exclusion limits obtained for the two channels
and for their combination.
Figure \ref{width} presents the \PZpr limits as a function of width instead of mass.

\begin{table}[htb]
\centering
\topcaption{Comparison of mass exclusion results (in \TeVns{}) for the individual channels and for their combination.}
\resizebox{\textwidth}{!}{
\begin{tabular}{lcccccccc}
       & \multicolumn{8}{c}{Excluded mass ranges [\TeVns{}]} \\
       & \multicolumn{2}{c}{\PZpr $(\Gamma/M=1\%)$} & \multicolumn{2}{c}{\PZpr $(\Gamma/M=10\%)$} & \multicolumn{2}{c}{\PZpr $(\Gamma/M=30\%)$} & \multicolumn{2}{c}{RS KK Gluon} \\
Result       &  Exp. & Obs. & Exp. & Obs. & Exp. & Obs. & Exp. & Obs. \\
\hline
Lepton+jets  & 0.6 -- 2.1 & 0.6 -- 2.3 & 0.5 -- 3.5 & 0.5 -- 3.4 & 0.5 -- 4.0 & 0.5 -- 4.0 & 0.5 -- 2.9 & 0.5 -- 2.9 \\
Fully hadronic  & 1.2 -- 1.8 & 1.4 -- 1.8 & 1.0 -- 3.2 & 1.0 -- 3.5 & 1.0 -- 3.7 & 1.0 -- 4.0 & 1.0 -- 2.6 & 1.0 -- 2.4 \\
Combined      & 0.6 -- 2.4 & 0.6 -- 2.5 & 0.5 -- 3.7 & 0.5 -- 3.9 & 0.5 -- 4.0 & 0.5 -- 4.0 & 0.5 -- 3.1 & 0.5 -- 3.3 \\
\end{tabular}
\label{compare_results}
}
\end{table}

\begin{table}[htb]
\centering
\topcaption{Expected and observed cross section limits at 95\% CL, for the 1\% width \PZpr resonance hypothesis.}
\begin{tabular}{ccccccc}
Mass [\TeVns{}] & Observed limits [pb] & \multicolumn{5}{c}{Expected limits [pb]} \\\cline{3-7}
 & & $-2\sigma$ & $-1\sigma$ & Median & $+1\sigma$ & $+2\sigma$ \\
\hline
0.5 & {78} & 32 & 50 & {88} & 150 & 230 \\
0.75 & {7.1} & 2.9 & 4.3 & {6.1} & 8.8 & 13 \\
1.0 & {1.8} & 0.75 & 1.0 & {1.5} & 2.2 & 3.0 \\
1.25 & {1.1} & 0.26 & 0.38 & {0.53} & 0.78 & 1.2 \\
1.5 & {0.24} & 0.15 & 0.20 & {0.29} & 0.43 & 0.62 \\
2.0 & {0.10} & 0.057 & 0.080 & {0.12} & 0.17 & 0.24 \\
2.5 & {0.046} & 0.031 & 0.044 & {0.061} & 0.090 & 0.13 \\
3.0 & {0.046} & 0.024 & 0.033 & {0.047} & 0.071 & 0.099 \\
3.5 & {0.025} & 0.019 & 0.026 & {0.036} & 0.055 & 0.081 \\
4.0 & {0.022} & 0.016 & 0.022 & {0.032} & 0.049 & 0.075 \\

\end{tabular}
\label{tab_zpn}

\end{table}

\begin{table}
\centering
\topcaption{Expected and observed cross section limits at 95\% CL, for the 10\% width \PZpr resonance hypothesis.}
\begin{tabular}{ccccccc}
Mass [\TeVns{}] & Observed limits [pb] & \multicolumn{5}{c}{Expected limits [pb]} \\\cline{3-7}
 & & $-2\sigma$ & $-1\sigma$ & Median & $+1\sigma$ & $+2\sigma$ \\
\hline
0.5 & {29} & 25 & 43 & {77} & 130 & 200 \\
0.75 & {9.1} & 3.9 & 5.6 & {8.1} & 12 & 18 \\
1.0 & {3.5} & 1.0 & 1.5 & {2.1} & 3.0 & 4.2 \\
1.25 & {1.3} & 0.41 & 0.55 & {0.79} & 1.2 & 1.6 \\
1.5 & {0.39} & 0.21 & 0.31 & {0.44} & 0.65 & 0.93 \\
2.0 & {0.15} & 0.089 & 0.12 & {0.18} & 0.25 & 0.38 \\
2.5 & {0.068} & 0.048 & 0.066 & {0.097} & 0.15 & 0.20 \\
3.0 & {0.067} & 0.039 & 0.053 & {0.074} & 0.11 & 0.16 \\
3.5 & {0.051} & 0.034 & 0.047 & {0.069} & 0.10 & 0.16 \\
4.0 & {0.050} & 0.035 & 0.048 & {0.072} & 0.11 & 0.17 \\

\end{tabular}
\label{tab_zpw}

\end{table}

\begin{table}
\centering
\topcaption{Expected and observed cross section limits at 95\% CL, for the 30\% width \PZpr resonance hypothesis.}
\begin{tabular}{ccccccc}
Mass [\TeVns{}] & Observed limits [pb] & \multicolumn{5}{c}{Expected limits [pb]} \\\cline{3-7}
 & & $-2\sigma$ & $-1\sigma$ & Median & $+1\sigma$ & $+2\sigma$ \\
\hline
1.0 & {3.6} & 1.2 & 1.6 & {2.4} & 3.7 & 5.3 \\
2.0 & {0.22} & 0.13 & 0.19 & {0.28} & 0.41 & 0.61 \\
3.0 & {0.12} & 0.080 & 0.11 & {0.15} & 0.22 & 0.32 \\
4.0 & {0.10} & 0.075 & 0.10 & {0.15} & 0.23 & 0.34 \\

\end{tabular}

\label{tab_zph}

\end{table}

\begin{table}
\centering
\topcaption{Expected and observed cross section limits at 95\% CL, for the RS KK gluon hypothesis.}
\begin{tabular}{ccccccc}
Mass [\TeVns{}] & Observed limits [pb] & \multicolumn{5}{c}{Expected limits [pb]} \\\cline{3-7}
 & & $-2\sigma$ & $-1\sigma$ & Median & $+1\sigma$ & $+2\sigma$ \\
\hline

0.5 & {41} & 26 & 40 & {69} & 130 & 190 \\
0.75 & {14} & 5.0 & 7.3 & {12} & 19 & 29 \\
1.0 & {4.4} & 1.3 & 1.7 & {2.5} & 3.8 & 5.3 \\
1.25 & {2.2} & 0.53 & 0.76 & {1.1} & 1.7 & 2.4 \\
1.5 & {0.73} & 0.33 & 0.44 & {0.67} & 1.0 & 1.4 \\
2.0 & {0.21} & 0.13 & 0.18 & {0.27} & 0.40 & 0.54 \\
2.5 & {0.13} & 0.082 & 0.12 & {0.17} & 0.25 & 0.37 \\
3.0 & {0.11} & 0.071 & 0.094 & {0.13} & 0.19 & 0.29 \\
3.5 & {0.093} & 0.065 & 0.088 & {0.13} & 0.20 & 0.29 \\
4.0 & {0.096} & 0.073 & 0.099 & {0.14} & 0.22 & 0.32 \\

\end{tabular}

\label{tab_kkg}

\end{table}

\section{Summary}
\label{sec:conclusions}

A model-independent search for the production of heavy
spin-1 or spin-2 resonances decaying into $\ttbar$ final states has been conducted. The data correspond to an integrated
luminosity of 2.6\fbinv collected with the CMS detector in
proton-proton collisions at $\sqrt{s} = 13\TeV$ at the LHC.
The analysis is designed to have high sensitivity at resonance masses above $1\TeV$,
where final-state decay products become collimated because of the large Lorentz boosts of the top quarks.
The analysis method provides an in-situ measurement of the data-to-simulation scale factor
for the $\ttagging$ efficiency and the normalization of the main backgrounds.
No evidence for massive resonances that decay to $\ttbar$ is found.
Limits at 95\% CL are set on the production cross section of new spin-1 particles
decaying to $\ttbar$ with relative decay widths
that are either narrow or wide compared with the detector resolution.

In addition, limits are set
on the production of particles in benchmark models beyond the standard model.
Topcolor \PZpr bosons with relative widths $\Gamma/M$ of
$1\%$, $10\%$, and $30\%$ are excluded for mass ranges of  $0.6$--$2.5$, $0.5$--$3.9$, and $0.5$--$4.0\TeV$, respectively.
Kaluza--Klein excitations of a gluon with masses in the range $0.5$--$3.3\TeV$
in the Randall--Sundrum model are also excluded.  This search presents limits on \PZpr bosons as a function of the relative width of the resonance in the range from 1--30\%, for the first time in CMS.

This analysis yields approximately the same sensitivity as the previous search based on
$8\TeV$ data~\cite{Khachatryan:2015sma} (corresponding to an integrated luminosity of 19.7\fbinv) for resonance masses in the range 1.0--2.0\TeV.
At higher resonance masses, the present analysis is significantly more
sensitive. Previous lower mass limits on the \PZpr with 10\% relative width and
the Kaluza--Klein gluon were
$2.9$ and $2.8\TeV$, respectively. The present analysis
extends the lower mass limits to 3.9 and 3.3\TeV, respectively, for these models.

\clearpage
\begin{acknowledgments}
\hyphenation{Bundes-ministerium Forschungs-gemeinschaft Forschungs-zentren} We congratulate our colleagues in the CERN accelerator departments for the excellent performance of the LHC and thank the technical and administrative staffs at CERN and at other CMS institutes for their contributions to the success of the CMS effort. In addition, we gratefully acknowledge the computing centers and personnel of the Worldwide LHC Computing Grid for delivering so effectively the computing infrastructure essential to our analyses. Finally, we acknowledge the enduring support for the construction and operation of the LHC and the CMS detector provided by the following funding agencies: the Austrian Federal Ministry of Science, Research and Economy and the Austrian Science Fund; the Belgian Fonds de la Recherche Scientifique, and Fonds voor Wetenschappelijk Onderzoek; the Brazilian Funding Agencies (CNPq, CAPES, FAPERJ, and FAPESP); the Bulgarian Ministry of Education and Science; CERN; the Chinese Academy of Sciences, Ministry of Science and Technology, and National Natural Science Foundation of China; the Colombian Funding Agency (COLCIENCIAS); the Croatian Ministry of Science, Education and Sport, and the Croatian Science Foundation; the Research Promotion Foundation, Cyprus; the Secretariat for Higher Education, Science, Technology and Innovation, Ecuador; the Ministry of Education and Research, Estonian Research Council via IUT23-4 and IUT23-6 and European Regional Development Fund, Estonia; the Academy of Finland, Finnish Ministry of Education and Culture, and Helsinki Institute of Physics; the Institut National de Physique Nucl\'eaire et de Physique des Particules~/~CNRS, and Commissariat \`a l'\'Energie Atomique et aux \'Energies Alternatives~/~CEA, France; the Bundesministerium f\"ur Bildung und Forschung, Deutsche Forschungsgemeinschaft, and Helmholtz-Gemeinschaft Deutscher Forschungszentren, Germany; the General Secretariat for Research and Technology, Greece; the National Scientific Research Foundation, and National Innovation Office, Hungary; the Department of Atomic Energy and the Department of Science and Technology, India; the Institute for Studies in Theoretical Physics and Mathematics, Iran; the Science Foundation, Ireland; the Istituto Nazionale di Fisica Nucleare, Italy; the Ministry of Science, ICT and Future Planning, and National Research Foundation (NRF), Republic of Korea; the Lithuanian Academy of Sciences; the Ministry of Education, and University of Malaya (Malaysia); the Mexican Funding Agencies (BUAP, CINVESTAV, CONACYT, LNS, SEP, and UASLP-FAI); the Ministry of Business, Innovation and Employment, New Zealand; the Pakistan Atomic Energy Commission; the Ministry of Science and Higher Education and the National Science Centre, Poland; the Funda\c{c}\~ao para a Ci\^encia e a Tecnologia, Portugal; JINR, Dubna; the Ministry of Education and Science of the Russian Federation, the Federal Agency of Atomic Energy of the Russian Federation, Russian Academy of Sciences, and the Russian Foundation for Basic Research; the Ministry of Education, Science and Technological Development of Serbia; the Secretar\'{\i}a de Estado de Investigaci\'on, Desarrollo e Innovaci\'on and Programa Consolider-Ingenio 2010, Spain; the Swiss Funding Agencies (ETH Board, ETH Zurich, PSI, SNF, UniZH, Canton Zurich, and SER); the Ministry of Science and Technology, Taipei; the Thailand Center of Excellence in Physics, the Institute for the Promotion of Teaching Science and Technology of Thailand, Special Task Force for Activating Research and the National Science and Technology Development Agency of Thailand; the Scientific and Technical Research Council of Turkey, and Turkish Atomic Energy Authority; the National Academy of Sciences of Ukraine, and State Fund for Fundamental Researches, Ukraine; the Science and Technology Facilities Council, UK; the US Department of Energy, and the US National Science Foundation.

Individuals have received support from the Marie-Curie programme and the European Research Council and EPLANET (European Union); the Leventis Foundation; the A. P. Sloan Foundation; the Alexander von Humboldt Foundation; the Belgian Federal Science Policy Office; the Fonds pour la Formation \`a la Recherche dans l'Industrie et dans l'Agriculture (FRIA-Belgium); the Agentschap voor Innovatie door Wetenschap en Technologie (IWT-Belgium); the Ministry of Education, Youth and Sports (MEYS) of the Czech Republic; the Council of Science and Industrial Research, India; the HOMING PLUS programme of the Foundation for Polish Science, cofinanced from European Union, Regional Development Fund, the Mobility Plus programme of the Ministry of Science and Higher Education, the National Science Center (Poland), contracts Harmonia 2014/14/M/ST2/00428, Opus 2013/11/B/ST2/04202, 2014/13/B/ST2/02543 and 2014/15/B/ST2/03998, Sonata-bis 2012/07/E/ST2/01406; the Thalis and Aristeia programmes cofinanced by EU-ESF and the Greek NSRF; the National Priorities Research Program by Qatar National Research Fund; the Programa Clar\'in-COFUND del Principado de Asturias; the Rachadapisek Sompot Fund for Postdoctoral Fellowship, Chulalongkorn University and the Chulalongkorn Academic into Its 2nd Century Project Advancement Project (Thailand); and the Welch Foundation, contract C-1845.
\end{acknowledgments}
\bibliography{auto_generated}

\providecommand{\href}[2]{#2}\begingroup\raggedright\begin{thebibliography}{10}%
\makeatletter
\providecommand{\hrefCMSnoop }[0]{\@secondoftwo}%
\makeatother
\providecommand{\doi}{\texttt{doi:}\begingroup \urlstyle{tt}\Url}

\bibitem{zprime_Rosner}
\hrefCMSnoop {}{J.~L. Rosner, ``{Prominent decay modes of a leptophobic
  $Z^\prime$}'',} \textit{ Phys. Lett. B} \textbf{ 387} (1996) 113,
  \href{http://dx.doi.org/10.1016/0370-2693(96)01022-2}{\doi{10.1016/0370-2693(96)01022-2}},
\href{http://www.arXiv.org/abs/hep-ph/9607207}{\texttt{arXiv:hep-ph/9607207}}.

\bibitem{zprime_Lynch}
\hrefCMSnoop {}{K.~R. Lynch, S.~Mrenna, M.~Narain, and E.~H. Simmons, ``Finding
  {$Z'$} bosons coupled preferentially to the third family at {CERN LEP} and
  the {Fermilab Tevatron}'',} \textit{ Phys. Rev. D} \textbf{ 63} (2001)
  035006,
  \href{http://dx.doi.org/10.1103/PhysRevD.63.035006}{\doi{10.1103/PhysRevD.63.035006}},
  \href{http://www.arXiv.org/abs/hep-ph/0007286}{\texttt{arXiv:hep-ph/0007286}}.

\bibitem{zprime_Carena}
\hrefCMSnoop {}{M.~Carena, A.~Daleo, B.~A. Dobrescu, and T.~M.~P. Tait,
  ``{$Z'$} gauge bosons at the {Fermilab Tevatron}'',} \textit{ Phys. Rev. D}
  \textbf{ 70} (2004) 093009,
  \href{http://dx.doi.org/10.1103/PhysRevD.70.093009}{\doi{10.1103/PhysRevD.70.093009}},
  \href{http://www.arXiv.org/abs/hep-ph/0408098}{\texttt{arXiv:hep-ph/0408098}}.

\bibitem{Hill1991419}
\hrefCMSnoop {}{C.~T. Hill, ``{Topcolor} top quark condensation in a gauge
  extension of the standard model'',} \textit{ Phys. Lett. B} \textbf{ 266}
  (1991) 419,
\href{http://dx.doi.org/10.1016/0370-2693(91)91061-Y}{\doi{10.1016/0370-2693(91)91061-Y}}.

\bibitem{Jain11124928}
\hrefCMSnoop {}{R.~M. Harris and S.~Jain, ``Cross sections for leptophobic
  topcolor $Z'$ decaying to top-antitop'',} \textit{ Eur. Phys. J. C} \textbf{
  72} (2012) 2072,
  \href{http://dx.doi.org/10.1140/epjc/s10052-012-2072-4}{\doi{10.1140/epjc/s10052-012-2072-4}},
  \href{http://www.arXiv.org/abs/1112.4928}{\texttt{arXiv:1112.4928}}.

\bibitem{Hill:1993hs}
\hrefCMSnoop {}{C.~T. Hill and S.~J. Parke, ``{Top production: Sensitivity to
  new physics}'',} \textit{ Phys. Rev. D} \textbf{ 49} (1994) 4454,
  \href{http://dx.doi.org/10.1103/PhysRevD.49.4454}{\doi{10.1103/PhysRevD.49.4454}},
\href{http://www.arXiv.org/abs/hep-ph/9312324}{\texttt{arXiv:hep-ph/9312324}}.

\bibitem{Hill:1994hp}
\hrefCMSnoop {}{C.~T. Hill, ``{Topcolor assisted technicolor}'',} \textit{
  Phys. Lett. B} \textbf{ 345} (1995) 483,
  \href{http://dx.doi.org/10.1016/0370-2693(94)01660-5}{\doi{10.1016/0370-2693(94)01660-5}},
\href{http://www.arXiv.org/abs/hep-ph/9411426}{\texttt{arXiv:hep-ph/9411426}}.

\bibitem{axigluon}
\hrefCMSnoop {}{P.~H. Frampton and S.~L. Glashow, ``Chiral color: An
  alternative to the standard model'',} \textit{ Phys. Lett. B} \textbf{ 190}
  (1987) 157,
  \href{http://dx.doi.org/10.1016/0370-2693(87)90859-8}{\doi{10.1016/0370-2693(87)90859-8}}.

\bibitem{Choudhury:2007ux}
\hrefCMSnoop {}{D.~Choudhury, R.~M. Godbole, R.~K. Singh, and K.~Wagh, ``Top
  production at the {Tevatron/LHC} and nonstandard, strongly interacting spin
  one particles'',} \textit{ Phys. Lett. B} \textbf{ 657} (2007) 69,
  \href{http://dx.doi.org/10.1016/j.physletb.2007.09.057}{\doi{10.1016/j.physletb.2007.09.057}},
\href{http://www.arXiv.org/abs/0705.1499}{\texttt{arXiv:0705.1499}}.

\bibitem{Godbole:2008qw}
\href {https://inspirehep.net/record/800069/files/arXiv:0810.3635.pdf}{R.~M.
  Godbole and D.~Choudhury, ``{Nonstandard, strongly interacting spin one $t
  \bar{t}$ resonances}'',} in \textit{ {Proceedings, 34th International
  Conference on High Energy Physics (ICHEP 2008): Philadelphia, Pennsylvania,
  July 30-August 5, 2008}}.
\newblock 2008.
\newblock
\href{http://www.arXiv.org/abs/0810.3635}{\texttt{arXiv:0810.3635}}.
\newblock

\bibitem{pseudohiggs}
\hrefCMSnoop {}{D.~Dicus, A.~Stange, and S.~Willenbrock, ``{Higgs} decay to top
  quarks at hadron colliders'',} \textit{ Phys. Lett. B} \textbf{ 333} (1994)
  126,
  \href{http://dx.doi.org/10.1016/0370-2693(94)91017-0}{\doi{10.1016/0370-2693(94)91017-0}},
  \href{http://www.arXiv.org/abs/hep-ph/9404359}{\texttt{arXiv:hep-ph/9404359}}.

\bibitem{Agashe:2006hk}
K.~Agashe\hrefCMSnoop {}{ {et~al.}, ``{CERN LHC} signals from warped extra
  dimensions'',} \textit{ Phys. Rev. D} \textbf{ 77} (2008) 015003,
  \href{http://dx.doi.org/10.1103/PhysRevD.77.015003}{\doi{10.1103/PhysRevD.77.015003}},
\href{http://www.arXiv.org/abs/hep-ph/0612015}{\texttt{arXiv:hep-ph/0612015}}.

\bibitem{Agashe:2007ki}
K.~Agashe\hrefCMSnoop {}{ {et~al.}, ``{CERN LHC} signals for warped electroweak
  neutral gauge bosons'',} \textit{ Phys. Rev. D} \textbf{ 76} (2007) 115015,
  \href{http://dx.doi.org/10.1103/PhysRevD.76.115015}{\doi{10.1103/PhysRevD.76.115015}},
\href{http://www.arXiv.org/abs/0709.0007}{\texttt{arXiv:0709.0007}}.

\bibitem{graviton}
\hrefCMSnoop {}{H.~Davoudiasl, J.~L. Hewett, and T.~G. Rizzo, ``Phenomenology
  of the {Randall-Sundrum} Gauge Hierarchy Model'',} \textit{ Phys. Rev. Lett.}
  \textbf{ 84} (2000) 2080,
  \href{http://dx.doi.org/10.1103/PhysRevLett.84.2080}{\doi{10.1103/PhysRevLett.84.2080}},
  \href{http://www.arXiv.org/abs/hep-ph/9909255}{\texttt{arXiv:hep-ph/9909255}}.

\bibitem{RandallSundrum}
\hrefCMSnoop {}{L.~Randall and R.~Sundrum, ``Large Mass Hierarchy from a Small
  Extra Dimension'',} \textit{ Phys. Rev. Lett.} \textbf{ 83} (1999) 3370,
  \href{http://dx.doi.org/10.1103/PhysRevLett.83.3370}{\doi{10.1103/PhysRevLett.83.3370}},
  \href{http://www.arXiv.org/abs/hep-ph/9905221}{\texttt{arXiv:hep-ph/9905221}}.

\bibitem{Randall:1999vf}
\hrefCMSnoop {}{L.~Randall and R.~Sundrum, ``An Alternative to
  Compactification'',} \textit{ Phys. Rev. Lett.} \textbf{ 83} (1999) 4690,
  \href{http://dx.doi.org/10.1103/PhysRevLett.83.4690}{\doi{10.1103/PhysRevLett.83.4690}},
\href{http://www.arXiv.org/abs/hep-th/9906064}{\texttt{arXiv:hep-th/9906064}}.

\bibitem{cdftt3}
\hrefCMSnoop {}{{CDF} Collaboration, ``A search for resonant production of $t
  \bar{t}$ pairs in $4.8\ \rm{fb}^{-1}$ of integrated luminosity of $p\bar{p}$
  collisions at {$\sqrt{s}=1.96\ \rm{TeV}$}'',} \textit{ Phys. Rev. D} \textbf{
  84} (2011) 072004,
  \href{http://dx.doi.org/10.1103/PhysRevD.84.072004}{\doi{10.1103/PhysRevD.84.072004}},
\href{http://www.arXiv.org/abs/1107.5063}{\texttt{arXiv:1107.5063}}.

\bibitem{d0_resonance}
\hrefCMSnoop {}{{D0} Collaboration, ``Search for a narrow $t \bar{t}$ resonance
  in ${p\bar{p}}$ collisions at {$\sqrt{s}=1.96$ TeV}'',} \textit{ Phys. Rev.
  D} \textbf{ 85} (2012) 051101,
  \href{http://dx.doi.org/10.1103/PhysRevD.85.051101}{\doi{10.1103/PhysRevD.85.051101}},
\href{http://www.arXiv.org/abs/1111.1271}{\texttt{arXiv:1111.1271}}.

\bibitem{cms-allhad}
\hrefCMSnoop {}{{CMS} Collaboration, ``{Search for anomalous $\ttbar$
  production in the highly-boosted all-hadronic final state}'',} \textit{ JHEP}
  \textbf{ 09} (2012) 029,
  \href{http://dx.doi.org/10.1007/JHEP09(2012)029}{\doi{10.1007/JHEP09(2012)029}},
\href{http://www.arXiv.org/abs/1204.2488}{\texttt{arXiv:1204.2488}}.

\bibitem{atlas-allhad}
\hrefCMSnoop {}{{ATLAS} Collaboration, ``Search for resonances decaying into
  top-quark pairs using fully hadronic decays in $pp$ collisions with {ATLAS}
  at {$\sqrt{s}=7$ TeV}'',} \textit{ JHEP} \textbf{ 01} (2013) 116,
  \href{http://dx.doi.org/10.1007/JHEP01(2013)116}{\doi{10.1007/JHEP01(2013)116}},
\href{http://www.arXiv.org/abs/1211.2202}{\texttt{arXiv:1211.2202}}.

\bibitem{atlas-ljets}
\hrefCMSnoop {}{{ATLAS} Collaboration, ``A search for $t \bar{t}$ resonances in
  the lepton plus jets final state with {ATLAS} using 4.7 fb$^{-1}$ of $pp$
  collisions at {$\sqrt{s} =7$ TeV}'',} \textit{ Phys. Rev. D} \textbf{ 88}
  (2013) 012004,
  \href{http://dx.doi.org/10.1103/PhysRevD.88.012004}{\doi{10.1103/PhysRevD.88.012004}},
\href{http://www.arXiv.org/abs/1305.2756}{\texttt{arXiv:1305.2756}}.

\bibitem{atlas-ljets-boosted}
\hrefCMSnoop {}{{ATLAS} Collaboration, ``A search for $t \bar{t}$ resonances in
  lepton+jets events with highly boosted top quarks collected in $pp$
  collisions at {$\sqrt{s} = 7$ TeV with the ATLAS detector}'',} \textit{ JHEP}
  \textbf{ 09} (2012) 041,
  \href{http://dx.doi.org/10.1007/JHEP09(2012)041}{\doi{10.1007/JHEP09(2012)041}},
\href{http://www.arXiv.org/abs/1207.2409}{\texttt{arXiv:1207.2409}}.

\bibitem{cms-ljets}
\hrefCMSnoop {}{{CMS} Collaboration, ``{Search for resonant $\ttbar$ production
  in lepton+jets events in pp collisions at $\sqrt{s}=7$ TeV}'',} \textit{
  JHEP} \textbf{ 12} (2012) 015,
  \href{http://dx.doi.org/10.1007/JHEP12(2012)015}{\doi{10.1007/JHEP12(2012)015}},
\href{http://www.arXiv.org/abs/1209.4397}{\texttt{arXiv:1209.4397}}.

\bibitem{Chatrchyan:2012yca}
\hrefCMSnoop {}{{CMS} Collaboration, ``{Search for $Z'$ resonances decaying to
  $t\bar{t}$ in dilepton+jets final states in pp collisions at $\sqrt{s}=7$
  TeV}'',} \textit{ Phys. Rev. D} \textbf{ 87} (2013) 072002,
  \href{http://dx.doi.org/10.1103/PhysRevD.87.072002}{\doi{10.1103/PhysRevD.87.072002}},
\href{http://www.arXiv.org/abs/1211.3338}{\texttt{arXiv:1211.3338}}.

\bibitem{Chatrchyan:2013lca}
\hrefCMSnoop {}{{CMS} Collaboration, ``Searches for new physics using the $t
  \bar{t}$ invariant mass distribution in $pp$ collisions at {$\sqrt{s} = 8$
  TeV}'',} \textit{ Phys. Rev. Lett.} \textbf{ 111} (2013) 211804,
  \href{http://dx.doi.org/10.1103/PhysRevLett.111.211804}{\doi{10.1103/PhysRevLett.111.211804}},
\href{http://www.arXiv.org/abs/1309.2030}{\texttt{arXiv:1309.2030}}.

\bibitem{Aad:2015fna}
\hrefCMSnoop {}{{ATLAS} Collaboration, ``{A search for $ \ttbar $ resonances
  using lepton-plus-jets events in proton-proton collisions at $\sqrt{s}=8 $
  TeV with the ATLAS detector}'',} \textit{ JHEP} \textbf{ 08} (2015) 148,
  \href{http://dx.doi.org/10.1007/JHEP08(2015)148}{\doi{10.1007/JHEP08(2015)148}},
\href{http://www.arXiv.org/abs/1505.07018}{\texttt{arXiv:1505.07018}}.

\bibitem{Khachatryan:2015sma}
\hrefCMSnoop {}{{CMS} Collaboration, ``{Search for resonant $\ttbar$ production
  in proton-proton collisions at $\sqrt{s}=8$~TeV}'',} \textit{ Phys. Rev. D}
  \textbf{ 93} (2016) 012001,
  \href{http://dx.doi.org/10.1103/PhysRevD.93.012001}{\doi{10.1103/PhysRevD.93.012001}},
\href{http://www.arXiv.org/abs/1506.03062}{\texttt{arXiv:1506.03062}}.

\bibitem{Altarelli:1989ff}
\hrefCMSnoop {}{G.~Altarelli, B.~Mele, and M.~Ruiz-Altaba, ``Searching for new
  heavy vector bosons in ${p\bar{p}}$ colliders'',} \textit{ Z. Phys. C}
  \textbf{ 45} (1989) 109,
  \href{http://dx.doi.org/10.1007/BF01556677}{\doi{10.1007/BF01556677}}.
[Erratum: \DOI{10.1007/BF01552335}].

\bibitem{Chatrchyan:2008zzk}
\hrefCMSnoop {}{{CMS} Collaboration, ``The {CMS} experiment at the {CERN}
  {LHC}'',} \textit{ JINST} \textbf{ 3} (2008) S08004,
\href{http://dx.doi.org/10.1088/1748-0221/3/08/S08004}{\doi{10.1088/1748-0221/3/08/S08004}}.

\bibitem{CMS-PAS-PFT-09-001}
\href {http://cdsweb.cern.ch/record/1194487}{{CMS} Collaboration,
  ``Particle--Flow Event Reconstruction in {CMS} and Performance for Jets,
  Taus, and {\MET}'',} CMS Physics Analysis Summary CMS-PAS-PFT-09-001, 2009.

\bibitem{CMS-PAS-PFT-10-001}
\href {http://cdsweb.cern.ch/record/1247373}{{CMS} Collaboration,
  ``Commissioning of the Particle-flow Event Reconstruction with the first
  {LHC} collisions recorded in the {CMS} detector'',} CMS Physics Analysis
  Summary CMS-PAS-PFT-10-001, 2010.

\bibitem{Chatrchyan:2014fea}
\hrefCMSnoop {}{{CMS} Collaboration, ``{Description and performance of track
  and primary-vertex reconstruction with the CMS tracker}'',} \textit{ JINST}
  \textbf{ 9} (2014) P10009,
  \href{http://dx.doi.org/10.1088/1748-0221/9/10/P10009}{\doi{10.1088/1748-0221/9/10/P10009}},
\href{http://www.arXiv.org/abs/1405.6569}{\texttt{arXiv:1405.6569}}.

\bibitem{muonreco}
\hrefCMSnoop {}{{CMS} Collaboration, ``{Performance of CMS muon reconstruction
  in pp collision events at $\sqrt{s}=7$ TeV}'',} \textit{ JINST} \textbf{ 7}
  (2012) P10002,
  \href{http://dx.doi.org/10.1088/1748-0221/7/10/P10002}{\doi{10.1088/1748-0221/7/10/P10002}},
\href{http://www.arXiv.org/abs/1206.4071}{\texttt{arXiv:1206.4071}}.

\bibitem{electronreco}
\hrefCMSnoop {}{{CMS} Collaboration, ``{Performance of electron reconstruction
  and selection with the CMS Detector in proton-proton collisions at
  $\sqrt{s}=8$~TeV}'',} \textit{ JINST} \textbf{ 10} (2015) P06005,
  \href{http://dx.doi.org/10.1088/1748-0221/10/06/P06005}{\doi{10.1088/1748-0221/10/06/P06005}},
\href{http://www.arXiv.org/abs/1502.02701}{\texttt{arXiv:1502.02701}}.

\bibitem{Cacciari:2008gp}
\hrefCMSnoop {}{M.~Cacciari, G.~P. Salam, and G.~Soyez, ``The anti-$k_t$ jet
  clustering algorithm'',} \textit{ JHEP} \textbf{ 04} (2008) 063,
  \href{http://dx.doi.org/10.1088/1126-6708/2008/04/063}{\doi{10.1088/1126-6708/2008/04/063}},
\href{http://www.arXiv.org/abs/0802.1189}{\texttt{arXiv:0802.1189}}.

\bibitem{FastJet}
\hrefCMSnoop {}{M.~Cacciari, G.~P. Salam, and G.~Soyez, ``{FastJet user
  manual}'',} \textit{ Eur. Phys. J. C} \textbf{ 72} (2012) 1896,
  \href{http://dx.doi.org/10.1140/epjc/s10052-012-1896-2}{\doi{10.1140/epjc/s10052-012-1896-2}},
\href{http://www.arXiv.org/abs/1111.6097}{\texttt{arXiv:1111.6097}}.

\bibitem{Cacciari:2008gn}
\hrefCMSnoop {}{M.~Cacciari, G.~P. Salam, and G.~Soyez, ``{The catchment area
  of jets}'',} \textit{ JHEP} \textbf{ 04} (2008) 005,
  \href{http://dx.doi.org/10.1088/1126-6708/2008/04/005}{\doi{10.1088/1126-6708/2008/04/005}},
\href{http://www.arXiv.org/abs/0802.1188}{\texttt{arXiv:0802.1188}}.

\bibitem{Chatrchyan:2011ds}
\hrefCMSnoop {}{{CMS} Collaboration, ``{Determination of jet energy calibration
  and transverse momentum resolution in CMS}'',} \textit{ JINST} \textbf{ 6}
  (2011) P11002,
  \href{http://dx.doi.org/10.1088/1748-0221/6/11/P11002}{\doi{10.1088/1748-0221/6/11/P11002}},
\href{http://www.arXiv.org/abs/1107.4277}{\texttt{arXiv:1107.4277}}.

\bibitem{Chatrchyan:2012jua}
\hrefCMSnoop {}{{CMS} Collaboration, ``{Identification of b-quark jets with the
  CMS experiment}'',} \textit{ JINST} \textbf{ 8} (2013) P04013,
  \href{http://dx.doi.org/10.1088/1748-0221/8/04/P04013}{\doi{10.1088/1748-0221/8/04/P04013}},
\href{http://www.arXiv.org/abs/1211.4462}{\texttt{arXiv:1211.4462}}.

\bibitem{CMS:BTV-15-001}
\href {http://cds.cern.ch/record/2138504}{{CMS} Collaboration,
  ``{Identification of b quark jets at the CMS experiment in the LHC Run 2}'',}
  CMS Physics Analysis Summary CMS-PAS-BTV-15-001, 2016.

\bibitem{JME-15-002}
\href {http://cds.cern.ch/record/2126325}{{CMS} Collaboration, ``Top Tagging
  with New Approaches'',} CMS Physics Analysis Summary CMS-PAS-JME-15-002,
  2016.

\bibitem{CAcambridge}
\hrefCMSnoop {}{Y.~L. Dokshitzer, G.~D. Leder, S.~Moretti, and B.~R. Webber,
  ``Better jet clustering algorithms'',} \textit{ JHEP} \textbf{ 08} (1997)
  001,
  \href{http://dx.doi.org/10.1088/1126-6708/1997/08/001}{\doi{10.1088/1126-6708/1997/08/001}},
\href{http://www.arXiv.org/abs/hep-ph/9707323}{\texttt{arXiv:hep-ph/9707323}}.

\bibitem{CAaachen}
\hrefCMSnoop {}{M.~Wobisch and T.~Wengler, ``{Hadronization corrections to jet
  cross sections in deep- inelastic scattering}'',} (1998).
\href{http://www.arXiv.org/abs/hep-ph/9907280}{\texttt{arXiv:hep-ph/9907280}}.

\bibitem{mmdt}
\hrefCMSnoop {}{M.~Dasgupta, A.~Fregoso, S.~Marzani, and G.~P. Salam,
  ``{Towards an understanding of jet substructure}'',} \textit{ JHEP} \textbf{
  09} (2013) 029,
  \href{http://dx.doi.org/10.1007/JHEP09(2013)029}{\doi{10.1007/JHEP09(2013)029}},
\href{http://www.arXiv.org/abs/1307.0007}{\texttt{arXiv:1307.0007}}.

\bibitem{CMS-JME-14-001}
\href {http://cds.cern.ch/record/1751454}{{CMS} Collaboration, ``{Study of
  Pileup Removal Algorithms for Jets}'',} CMS Physics Analysis Summary
  CMS-PAS-JME-14-001, 2014.

\bibitem{Larkoski:2014wba}
\hrefCMSnoop {}{A.~J. Larkoski, S.~Marzani, G.~Soyez, and J.~Thaler, ``Soft
  drop'',} \textit{ JHEP} \textbf{ 05} (2014) 146,
  \href{http://dx.doi.org/10.1007/JHEP05(2014)146}{\doi{10.1007/JHEP05(2014)146}},
\href{http://www.arXiv.org/abs/1402.2657}{\texttt{arXiv:1402.2657}}.

\bibitem{Thaler:2010tr}
\hrefCMSnoop {}{J.~Thaler and K.~Van~Tilburg, ``Identifying boosted objects
  with {N-subjettiness}'',} \textit{ JHEP} \textbf{ 03} (2011) 015,
  \href{http://dx.doi.org/10.1007/JHEP03(2011)015}{\doi{10.1007/JHEP03(2011)015}},
\href{http://www.arXiv.org/abs/1011.2268}{\texttt{arXiv:1011.2268}}.

\bibitem{Thaler:2011gf}
\hrefCMSnoop {}{J.~Thaler and K.~Van~Tilburg, ``Maximizing boosted top
  identification by minimizing {$N$}-subjettiness'',} \textit{ JHEP} \textbf{
  02} (2012) 093,
  \href{http://dx.doi.org/10.1007/JHEP02(2012)093}{\doi{10.1007/JHEP02(2012)093}},
\href{http://www.arXiv.org/abs/1108.2701}{\texttt{arXiv:1108.2701}}.

\bibitem{Alwall:2014hca}
J.~Alwall\hrefCMSnoop {}{ {et~al.}, ``{The automated computation of tree-level
  and next-to-leading order differential cross sections, and their matching to
  parton shower simulations}'',} \textit{ JHEP} \textbf{ 07} (2014) 079,
  \href{http://dx.doi.org/10.1007/JHEP07(2014)079}{\doi{10.1007/JHEP07(2014)079}},
\href{http://www.arXiv.org/abs/1405.0301}{\texttt{arXiv:1405.0301}}.

\bibitem{Sjostrand:2006za}
\hrefCMSnoop {}{T.~Sj{\"o}strand, S.~Mrenna, and P.~Z. Skands, ``{PYTHIA 6.4}
  physics and manual'',} \textit{ JHEP} \textbf{ 05} (2006) 026,
  \href{http://dx.doi.org/10.1088/1126-6708/2006/05/026}{\doi{10.1088/1126-6708/2006/05/026}},
\href{http://www.arXiv.org/abs/hep-ph/0603175}{\texttt{arXiv:hep-ph/0603175}}.

\bibitem{Sjostrand:2014zea}
T.~Sj{\"o}strand\hrefCMSnoop {}{ {et~al.}, ``An introduction to {PYTHIA
  8.2}'',} \textit{ Comput. Phys. Commun.} \textbf{ 191} (2015) 159,
  \href{http://dx.doi.org/10.1016/j.cpc.2015.01.024}{\doi{10.1016/j.cpc.2015.01.024}},
\href{http://www.arXiv.org/abs/1410.3012}{\texttt{arXiv:1410.3012}}.

\bibitem{mlm}
\hrefCMSnoop {}{M.~L. Mangano, M.~Moretti, F.~Piccinini, and M.~Treccani,
  ``Matching matrix elements and shower evolution for top-quark production in
  hadronic collisions'',} \textit{ JHEP} \textbf{ 01} (2007) 013,
  \href{http://dx.doi.org/10.1088/1126-6708/2007/01/013}{\doi{10.1088/1126-6708/2007/01/013}},
  \href{http://www.arXiv.org/abs/hep-ph/0611129}{\texttt{arXiv:hep-ph/0611129}}.

\bibitem{Nason:2004rx}
\hrefCMSnoop {}{P.~Nason, ``{A new method for combining NLO QCD with shower
  Monte Carlo algorithms}'',} \textit{ JHEP} \textbf{ 11} (2004) 040,
  \href{http://dx.doi.org/10.1088/1126-6708/2004/11/040}{\doi{10.1088/1126-6708/2004/11/040}},
\href{http://www.arXiv.org/abs/hep-ph/0409146}{\texttt{arXiv:hep-ph/0409146}}.

\bibitem{Frixione:2007vw}
\hrefCMSnoop {}{S.~Frixione, P.~Nason, and C.~Oleari, ``{Matching NLO QCD
  computations with Parton Shower simulations: the POWHEG method}'',} \textit{
  JHEP} \textbf{ 11} (2007) 070,
  \href{http://dx.doi.org/10.1088/1126-6708/2007/11/070}{\doi{10.1088/1126-6708/2007/11/070}},
\href{http://www.arXiv.org/abs/0709.2092}{\texttt{arXiv:0709.2092}}.

\bibitem{Alioli:2010xd}
\hrefCMSnoop {}{S.~Alioli, P.~Nason, C.~Oleari, and E.~Re, ``{A general
  framework for implementing NLO calculations in shower Monte Carlo programs:
  the POWHEG BOX}'',} \textit{ JHEP} \textbf{ 06} (2010) 043,
  \href{http://dx.doi.org/10.1007/JHEP06(2010)043}{\doi{10.1007/JHEP06(2010)043}},
\href{http://www.arXiv.org/abs/1002.2581}{\texttt{arXiv:1002.2581}}.

\bibitem{Frixione:2007nw}
\hrefCMSnoop {}{S.~Frixione, P.~Nason, and G.~Ridolfi, ``{A positive-weight
  next-to-leading-order Monte Carlo for heavy flavour hadroproduction}'',}
  \textit{ JHEP} \textbf{ 09} (2007) 126,
  \href{http://dx.doi.org/10.1088/1126-6708/2007/09/126}{\doi{10.1088/1126-6708/2007/09/126}},
\href{http://www.arXiv.org/abs/0707.3088}{\texttt{arXiv:0707.3088}}.

\bibitem{Re:2010bp}
\hrefCMSnoop {}{E.~Re, ``{Single-top Wt-channel production matched with parton
  showers using the POWHEG method}'',} \textit{ Eur. Phys. J. C} \textbf{ 71}
  (2011) 1547,
  \href{http://dx.doi.org/10.1140/epjc/s10052-011-1547-z}{\doi{10.1140/epjc/s10052-011-1547-z}},
\href{http://www.arXiv.org/abs/1009.2450}{\texttt{arXiv:1009.2450}}.

\bibitem{Czakon:2011xx}
\hrefCMSnoop {}{M.~Czakon and A.~Mitov, ``{Top++: A program for the calculation
  of the top-pair cross-section at hadron colliders}'',} \textit{ Comput. Phys.
  Commun.} \textbf{ 185} (2014) 2930,
  \href{http://dx.doi.org/10.1016/j.cpc.2014.06.021}{\doi{10.1016/j.cpc.2014.06.021}},
\href{http://www.arXiv.org/abs/1112.5675}{\texttt{arXiv:1112.5675}}.

\bibitem{Li:2012wna}
\hrefCMSnoop {}{Y.~Li and F.~Petriello, ``{Combining QCD and electroweak
  corrections to dilepton production in FEWZ}'',} \textit{ Phys. Rev. D}
  \textbf{ 86} (2012) 094034,
  \href{http://dx.doi.org/10.1103/PhysRevD.86.094034}{\doi{10.1103/PhysRevD.86.094034}},
\href{http://www.arXiv.org/abs/1208.5967}{\texttt{arXiv:1208.5967}}.

\bibitem{Kant:2014oha}
P.~Kant\hrefCMSnoop {}{ {et~al.}, ``{HatHor for single top-quark production:
  Updated predictions and uncertainty estimates for single top-quark production
  in hadronic collisions}'',} \textit{ Comput. Phys. Commun.} \textbf{ 191}
  (2015) 74,
  \href{http://dx.doi.org/10.1016/j.cpc.2015.02.001}{\doi{10.1016/j.cpc.2015.02.001}},
\href{http://www.arXiv.org/abs/1406.4403}{\texttt{arXiv:1406.4403}}.

\bibitem{Kidonakis:2012rm}
\hrefCMSnoop {}{N.~Kidonakis, ``{NNLL threshold resummation for top-pair and
  single-top production}'',} \textit{ Phys. Part. Nucl.} \textbf{ 45} (2014)
  714,
  \href{http://dx.doi.org/10.1134/S1063779614040091}{\doi{10.1134/S1063779614040091}},
\href{http://www.arXiv.org/abs/1210.7813}{\texttt{arXiv:1210.7813}}.

\bibitem{Campbell:2010ff}
\hrefCMSnoop {}{J.~M. Campbell and R.~K. Ellis, ``{MCFM for the Tevatron and
  the LHC}'',} \textit{ Nucl. Phys. Proc. Suppl.} \textbf{ 205} (2010) 10,
  \href{http://dx.doi.org/10.1016/j.nuclphysbps.2010.08.011}{\doi{10.1016/j.nuclphysbps.2010.08.011}},
\href{http://www.arXiv.org/abs/1007.3492}{\texttt{arXiv:1007.3492}}.

\bibitem{Ball:2014uwa}
\hrefCMSnoop {}{{NNPDF} Collaboration, ``{Parton distributions for the LHC Run
  II}'',} \textit{ JHEP} \textbf{ 04} (2015) 040,
  \href{http://dx.doi.org/10.1007/JHEP04(2015)040}{\doi{10.1007/JHEP04(2015)040}},
\href{http://www.arXiv.org/abs/1410.8849}{\texttt{arXiv:1410.8849}}.

\bibitem{CMS-PAS-GEN-14-001}
\hrefCMSnoop {}{{CMS} Collaboration, ``{Event generator tunes obtained from
  underlying event and multiparton scattering measurements}'',} \textit{ Eur.
  Phys. J. C} \textbf{ 76} (2016) 155,
  \href{http://dx.doi.org/10.1140/epjc/s10052-016-3988-x}{\doi{10.1140/epjc/s10052-016-3988-x}},
\href{http://www.arXiv.org/abs/1512.00815}{\texttt{arXiv:1512.00815}}.

\bibitem{Skands:2014pea}
\hrefCMSnoop {}{P.~Skands, S.~Carrazza, and J.~Rojo, ``{Tuning PYTHIA 8.1: the
  Monash 2013 Tune}'',} \textit{ Eur. Phys. J. C} \textbf{ 74} (2014) 3024,
  \href{http://dx.doi.org/10.1140/epjc/s10052-014-3024-y}{\doi{10.1140/epjc/s10052-014-3024-y}},
\href{http://www.arXiv.org/abs/1404.5630}{\texttt{arXiv:1404.5630}}.

\bibitem{theta}
\href
  {www-ekp.physik.uni-karlsruhe.de/~ott/theta/testing/html/theta__auto__intro.html}{T.~M{\"u}ller,
  J.~Ott, and J.~Wagner-Kuhr, ``theta - a framework for template-based modeling
  and inference'',} 2010.

\bibitem{Campbell:2011bn}
\hrefCMSnoop {}{J.~M. Campbell, R.~K. Ellis, and C.~Williams, ``{Vector boson
  pair production at the LHC}'',} \textit{ JHEP} \textbf{ 07} (2011) 018,
  \href{http://dx.doi.org/10.1007/JHEP07(2011)018}{\doi{10.1007/JHEP07(2011)018}},
\href{http://www.arXiv.org/abs/1105.0020}{\texttt{arXiv:1105.0020}}.

\bibitem{Gehrmann:2014fva}
T.~Gehrmann\hrefCMSnoop {}{ {et~al.}, ``{$W^+W^-$ Production at Hadron
  Colliders in Next-to-Next-to-Leading-Order QCD}'',} \textit{ Phys. Rev.
  Lett.} \textbf{ 113} (2014) 212001,
  \href{http://dx.doi.org/10.1103/PhysRevLett.113.212001}{\doi{10.1103/PhysRevLett.113.212001}},
\href{http://www.arXiv.org/abs/1408.5243}{\texttt{arXiv:1408.5243}}.

\bibitem{Kidonakis:2013zqa}
\hrefCMSnoop {}{N.~Kidonakis, ``{Top Quark Production}'',} \textit{
  {Proceedings, Helmholtz International Summer School on Physics of Heavy
  Quarks and Hadrons (HQ 2013)}} (2014) 139,
  \href{http://dx.doi.org/10.3204/DESY-PROC-2013-03/Kidonakis}{\doi{10.3204/DESY-PROC-2013-03/Kidonakis}},
\href{http://www.arXiv.org/abs/1311.0283}{\texttt{arXiv:1311.0283}}.

\bibitem{Aliev:2010zk}
M.~Aliev\hrefCMSnoop {}{ {et~al.}, ``{HATHOR: HAdronic Top and Heavy quarks
  crOss section calculatoR}'',} \textit{ Comput. Phys. Commun.} \textbf{ 182}
  (2011) 1034,
  \href{http://dx.doi.org/10.1016/j.cpc.2010.12.040}{\doi{10.1016/j.cpc.2010.12.040}},
\href{http://www.arXiv.org/abs/1007.1327}{\texttt{arXiv:1007.1327}}.

\bibitem{Botje:2011sn}
M.~Botje\hrefCMSnoop {}{ {et~al.}, ``{The PDF4LHC Working Group Interim
  Recommendations}'',} (2011).
\href{http://www.arXiv.org/abs/1101.0538}{\texttt{arXiv:1101.0538}}.

\bibitem{CMS:LUM-15-001}
\href {http://cds.cern.ch/record/2138682}{{CMS} Collaboration, ``{CMS}
  Luminosity Measurement for the 2015 Data Taking Period'',} CMS Physics
  Analysis Summary CMS-PAS-LUM-15-001, 2016.

\bibitem{CMS-PAS-FSQ-15-005}
\href {https://cds.cern.ch/record/2145896}{{CMS} Collaboration, ``{Measurement
  of the inelastic proton-proton cross section at
  $\sqrt{s}=13~\mathrm{TeV}$}'',} CMS Physics Analysis Summary
  CMS-PAS-FSQ-15-005, 2016.

\bibitem{bayesbook}
A.~O'Hagan and J.~J. Forster, ``{Kendall's Advanced Theory of Statistics. Vol.
  2B: Bayesian Inference}''.
\newblock Arnold, London, 2004.
\newblock
  \href{http://dx.doi.org/10.1111/j.1467-985X.2004.00347_15.x}{\doi{10.1111/j.1467-985X.2004.00347_15.x}}.

\bibitem{barlow_beeston}
\hrefCMSnoop {}{R.~J. Barlow and C.~Beeston, ``{Fitting using finite Monte
  Carlo samples}'',} \textit{ Comput. Phys. Commun.} \textbf{ 77} (1993) 219,
\href{http://dx.doi.org/10.1016/0010-4655(93)90005-W}{\doi{10.1016/0010-4655(93)90005-W}}.

\bibitem{Bonciani:2015hgv}
R.~Bonciani\hrefCMSnoop {}{ {et~al.}, ``{Electroweak top-quark pair production
  at the LHC with $Z'$ bosons to NLO QCD in POWHEG}'',} \textit{ JHEP} \textbf{
  02} (2016) 141,
  \href{http://dx.doi.org/10.1007/JHEP02(2016)141}{\doi{10.1007/JHEP02(2016)141}},
\href{http://www.arXiv.org/abs/1511.08185}{\texttt{arXiv:1511.08185}}.

\bibitem{Gao:2010bb}
J.~Gao\hrefCMSnoop {}{ {et~al.}, ``{Next-to-leading order QCD corrections to
  the heavy resonance production and decay into top quark pair at the LHC}'',}
  \textit{ Phys. Rev. D} \textbf{ 82} (2010) 014020,
  \href{http://dx.doi.org/10.1103/PhysRevD.82.014020}{\doi{10.1103/PhysRevD.82.014020}},
\href{http://www.arXiv.org/abs/1004.0876}{\texttt{arXiv:1004.0876}}.

\end{thebibliography}\endgroup

\cleardoublepage \appendix\section{The CMS Collaboration \label{app:collab}}\begin{sloppypar}\hyphenpenalty=5000\widowpenalty=500\clubpenalty=5000\textbf{Yerevan Physics Institute,  Yerevan,  Armenia}\\*[0pt]
A.M.~Sirunyan, A.~Tumasyan
\vskip\cmsinstskip
\textbf{Institut f\"{u}r Hochenergiephysik,  Wien,  Austria}\\*[0pt]
W.~Adam, E.~Asilar, T.~Bergauer, J.~Brandstetter, E.~Brondolin, M.~Dragicevic, J.~Er\"{o}, M.~Flechl, M.~Friedl, R.~Fr\"{u}hwirth\cmsAuthorMark{1}, V.M.~Ghete, C.~Hartl, N.~H\"{o}rmann, J.~Hrubec, M.~Jeitler\cmsAuthorMark{1}, A.~K\"{o}nig, I.~Kr\"{a}tschmer, D.~Liko, T.~Matsushita, I.~Mikulec, D.~Rabady, N.~Rad, B.~Rahbaran, H.~Rohringer, J.~Schieck\cmsAuthorMark{1}, J.~Strauss, W.~Waltenberger, C.-E.~Wulz\cmsAuthorMark{1}
% ...
\vskip\cmsinstskip
1:~~Also at Vienna University of Technology, Vienna, Austria\\
% ...

\end{sloppypar}
\end{document}